\documentclass{pasa}%

\title[Feedback and feeding with {\it SPICA}]{Feedback and feeding in the
  context of galaxy evolution with {\it SPICA}: direct characterization of
  molecular outflows and inflows}
\author[Gonz\'alez-Alfonso et al.]{E. Gonz\'alez-Alfonso$^1$, 
L. Armus$^2$, 
F. J. Carrera$^3$,
V. Charmandaris$^4$,
A. Efstathiou$^5$,
E. Egami$^6$,
J. A. Fern\'andez-Ontiveros$^{7,8,9}$,
J. Fischer$^{10}$,
G. L. Granato$^{11}$,
C. Gruppioni$^{12}$,
E. Hatziminaoglou$^{13}$,
M. Imanishi$^{14}$,
N. Isobe$^{15}$,
H. Kaneda$^{16}$,
D. Koziel-Wierzbowska$^{17}$,
M. A. Malkan$^{18}$,
J. Mart\'{\i}n-Pintado$^{19}$,
S. Mateos$^{3}$,
H. Matsuhara$^{15}$,
G. Miniutti$^{20}$,
T. Nakagawa$^{15}$,
F. Pozzi$^{21,12}$,
F. Rico-Villas$^{19}$,
G. Rodighiero$^{22}$,
P. Roelfsema$^{23}$,
L. Spinoglio$^{9}$,
H. W. W. Spoon$^{24}$,
E. Sturm$^{25}$,
F. van der Tak$^{26}$,
C. Vignali$^{21,12}$,
\and L. Wang$^{26,27}$
\\
\affil{$^1$Universidad de Alcal\'a, Departamento de F\'{\i}sica
     y Matem\'aticas, Campus Universitario, E-28871 Alcal\'a de Henares,
     Madrid, Spain}
\affil{$^2$ IPAC, California Institute of Technology, Pasadena, CA 91125, USA}
\affil{$^3$ Instituto de F\'{\i}sica de Cantabria (CSIC-UC), Avenida de los
  Castros, E-39005 Santander, Spain}
\affil{$^4$ Institute for Astronomy, Astrophysics, Space Applications \&
  Remote Sensing, National Observatory of Athens, Athens, Greece} 
\affil{$^5$ School of Sciences, European University Cyprus, Diogenes Street,
  Engomi, 1516, Nicosia, Cyprus}
\affil{$^6$ Steward Observatory, University of Arizona, 933 North Cherry
  Avenue, Tucson, AZ 85721, USA} 
\affil{$^7$Instituto de Astrof\'{\i}sica de Canarias (IAC), E-38205 La Laguna,
Tenerife, Spain}
\affil{$^8$Universidad de La Laguna (ULL), Dpto. Astrof\'isica, E-38206 La
  Laguna, Tenerife, Spain} 
\affil{$^9$Istituto di Astrofisica e Planetologia Spaziali (INAF-IAPS), Via
  Fosso del Cavaliere 100, I-00133 Roma, Italy} 
\affil{$^{10}$ Naval Research Laboratory, Remote Sensing Division, 4555
     Overlook Ave SW, Washington, DC 20375, USA}
\affil{$^{11}$ INAF - Osservatorio Astronomico di Trieste, via Tiepolo 11,
  34131 Trieste, Italy} 
\affil{$^{12}$ INAF - Osservatorio Astronomico di Bologna, Via Gobetti 93/3,
  40129 Bologna, Italy}
\affil{$^{13}$ European Southern Observatory, Karl-Schwarzschild-Str. 2,
  D-85748 Garching bei M\"unchen, Germany} 
\affil{$^{14}$ National Astronomical Observatory of Japan, National Institutes
  of Natural Sciences (NINS), 2-21-1 Osawa, Mitaka, Tokyo, Japan} 
\affil{$^{15}$ Institute of Space and
  Astronautical Science (ISAS), JAXA, 3-1-1 Yoshinodai, Chuo-ku, Sagamihara,
  Kanagawa 252-5210, Japan} 
\affil{$^{16}$ Graduate School of Science, Nagoya University, Furo-cho,
  Chikusa-ku, Nagoya 464-8602, Japan} 
\affil{$^{17}$ Astronomical Observatory, Jagiellonian University, ul. Orla,
  PL-30-244 Krak\'ow, Poland} 
\affil{$^{18}$ Department of Physics and Astronomy, University of California,
  Los Angeles, CA, 90024, USA} 
\affil{$^{19}$ Centro de Astrobiolog\'{\i}a (CSIC-INTA), Ctra. de Torrej\'on a
  Ajalvir km 4, E-28850, Torrej\'on de Ardoz, Madrid, Spain} 
\affil{$^{20}$ Centro de Astrobiolog\'{i}a (CSIC--INTA), Depto. de
  Astrof\'{i}sica, ESAC campus, Camino Bajo del Castillo s/n, E-28692
  Villanueva de la Ca\~{n}ada, Spain}
\affil{$^{21}$ Dipartimento di Fisica e Astronomia, Alma Mater Studiorum,
  Universit\`a degli Studi di Bologna, Via Gobetti 93/2, 40129 Bologna, 
  Italy}   
\affil{$^{22}$ Dipartimento di Fisica e Astronomia, Universit\`a di Padova,
  vicolo dell'Osservatorio 2, 35122, Padova, Italy} 
\affil{$^{23}$ SRON Netherlands Institute for Space Research, Postbus 800,
  9700, AV Groningen, The Netherlands} 
\affil{$^{24}$ Cornell University, Cornell Center for Astrophysics and
  Planetary Science, Ithaca, NY 14853, USA}
\affil{$^{25}$ Max-Planck-Institute for Extraterrestrial Physics (MPE),
  Giessenbachstra{\ss}e 1, 85748 Garching, Germany}
\affil{$^{26}$ SRON Netherlands Institute for Space Research, Landleven 12,
  9747 AD, Groningen, The Netherlands} 
\affil{$^{27}$ Kapteyn Astronomical Institute, University of Groningen,
  Postbus 800, 9700 AV Groningen, the Netherlands} 
}%
\jid{PASA}
\doi{10.1017/pas.\the\year.xxx}
\jyear{\the\year}

\usepackage[authoryear]{natbib}
\usepackage{graphicx}
\bibpunct{(}{)}{;}{a}{}{,}
\usepackage{aas_macros}
\usepackage{hyperref} 
\hypersetup{colorlinks,citecolor=blue,linkcolor=blue,urlcolor=blue}

\newcommand{\kms}{{\hbox {km\thinspace s$^{-1}$}}}
\newcommand{\Lsun}{{\hbox {L$_\odot$}}}
\newcommand{\Msun}{{\hbox {M$_\odot$}}}

\newcommand{\hdo}{{\hbox {H$_{2}$O}}}

\newcommand{\tdust}{{\hbox {$T_{\mathrm{dust}}$}}}

\begin{document}%
\begin{abstract}
A far-infrared observatory such as the {\it SPace Infrared telescope for
  Cosmology and Astrophysics} ({\it SPICA}), with its unprecedented
 spectroscopic sensitivity, 
would unveil the role of feedback in galaxy evolution during the last $\sim10$
Gyr of the Universe ($z=1.5-2$), 
 through the use of far- and mid-infrared molecular and ionic fine
  structure lines that trace outflowing and infalling gas.
Outflowing gas is identified in the far-infrared through P-Cygni line shapes
and absorption blueshifted wings in molecular lines with high dipolar moments,
and through emission line wings of fine-structure lines of ionized gas.
We quantify the
detectability of galaxy-scale massive molecular and ionized outflows as a
function of redshift in AGN-dominated, starburst-dominated, and
main-sequence galaxies, explore the detectability of metal-rich inflows in the
local Universe, and describe the most significant synergies with other
current and future observatories that will measure feedback in galaxies
  via complementary tracers at other wavelengths.
\end{abstract}
\maketitle%

{\bf Preface}

\vspace{0.5cm}
\noindent
The following set of papers describe in detail the science goals of the future
Space Infrared telescope for Cosmology and Astrophysics (SPICA). The SPICA
satellite will employ a 2.5-m telescope, actively cooled to around 6\,K, and a
suite of mid- to far-IR spectrometers and photometric cameras, equipped with
state of the art detectors. In particular the SPICA Far Infrared Instrument
(SAFARI) will be a grating spectrograph with low (R=300) and medium
(R$\simeq$3000--11000) resolution observing modes instantaneously covering the
35--230\,$\mu$m wavelength range. The SPICA Mid-Infrared Instrument (SMI) will
have three operating modes:  a large field of view (12'$\times$10')
low-resolution 17--36\,$\mu$m spectroscopic (R$\sim$50--120) and photometric
camera at 34$\mu$m, a medium resolution (R$\simeq$2000)  grating spectrometer
covering wavelengths of 17--36\,$\mu$m and a high-resolution echelle module
(R$\simeq$28000) for the 12--18\,$\mu$m domain.  A  large field of view
(80''$\times$80''), three channel, (110\,$\mu$m, 220\,$\mu$m and 350\,$\mu$m)
polarimetric camera will also be part of the instrument complement. These
articles will focus on some of the major scientific questions that the SPICA
mission aims to address, more details about the mission and instruments can be
found in \citet{roe17}.

\section{INTRODUCTION }
\label{sec:intro}

The tight 
correlations found between the masses of supermassive black holes (SMBH) and
the velocity dispersion, the stellar mass, and luminosity of the spheroidal
components of their host galaxies 
\citep[e.g.][]{mag98,fer00,tre02} suggest a link between
the growth of the BH and galaxy formation/evolution. 
In addition, the bimodality of the color distribution of
local galaxies \citep[e.g.][]{str01,bal04}, with the blue
galaxies actively forming stars and red-and-dead galaxies evolving passively
from the earlier epochs of peak star formation ($z\sim2$), strongly suggests
that the color must have evolved rapidly, 
with star formation terminated on short timescales 
\citep[e.g.][]{hop06a,sch14}. These observations are consistent with a
self-regulated feedback model in which funneling of 
large amounts of gas into the nuclear regions of galaxies generate both
a nuclear starburst (SB) and the growth of a SMBH. Once the SMBH reaches a
threshold in mass/luminosity, the energy and momentum released by the
accreting SMBH couples with the surrounding interstellar medium (ISM),
limiting the accretion onto the SMBH and quenching the SBs via injection of
turbulence or through a fast
sweeping out of the ISM gas from which stars are formed (negative feedback), 
ultimately yielding the $M_{\mathrm{BH}}-\sigma$ relationship 
\citep{sil98,mat05,spr05,mur05,hop06b}.

\begin{figure*}
\begin{center}
\begin{tabular}{@{}cc@{}}
\includegraphics[width=8.2cm]{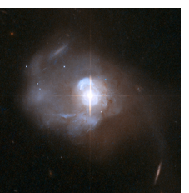}
& \includegraphics[width=8.25cm]{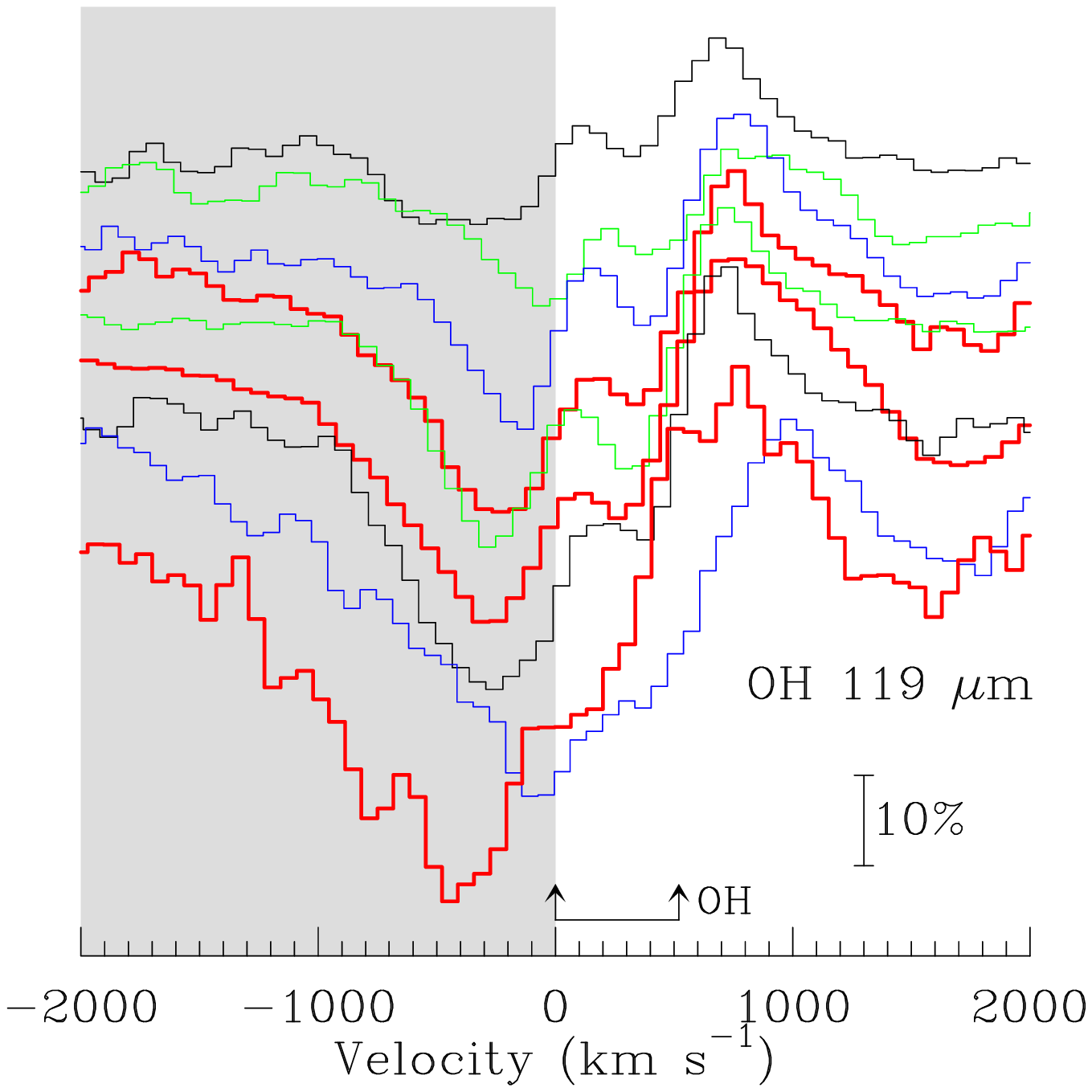} \
\end{tabular}
\caption{{\it Left}: Image of the nearest QSO Mrk 231 taken by the 
{\it Hubble Space Telescope} 
[Credit: NASA, ESA, the Hubble Heritage Team (STScI/AURA)- ESA/Hubble
 Collaboration, and A. Evans (University of Virginia,
 Charlottesville/NRAO/Stony Brook University)]. Mrk~231 shows evidence for
  powerful outflows at basically all wavelengths and ISM phases, with the
  molecular phase dominating the energetics (mass, momentum, and energy
  rates). {\it Right}: the OH doublet at 119 $\mu$m in 9 local
  ULIRGs observed with {\it Herschel}/PACS, selected as examples of ULIRGs
  with P-Cygni profiles to illustrate that the molecular gas is outflowing at
  high velocities ($800-1500$ \kms) in many such local gas-rich mergers. The
  three thick profiles in red correspond to the three local templates we use
  in this paper: IRAS~23365+3604, Mrk~231, and IRAS~03158+4227 (from top to
  bottom). 
  While the spectral resolution of {\it Herschel}/PACS at 119 $\mu$m is
  $R\approx1000$, {\it SPICA}/SAFARI will attain an even better resolution in
  its high resolution mode.
}
\label{art}
\end{center}
\end{figure*}

In the local Universe, evidence for galaxy-scale feedback is observed in all
phases of the interstellar medium (ISM) in luminous systems, and is now
established as a key ingredient of galaxy evolution. The most extreme cases of
feedback are observed in luminous and ultraluminous infrared galaxies 
((U)LIRGs, with $L_{\mathrm{IR}}>10^{11}$ and $>10^{12}$ \Lsun, 
respectively) and SBs, where superwinds observed in lines of ions and
neutral atoms are common 
\citep[e.g.][]{hec90,hec15,lip05,vei05,rup05,mar06,spo09,rod13,jan16}. The 
molecular phase is also crucial to understand feedback, because stars are
formed from molecular gas and indeed most gas mass in the central regions of 
gas rich/obscured galaxies is in molecular form and the column densities
  associated with the molecular gas are the highest.
Far-infrared (far-IR) spectroscopic observations by  
{\it Herschel}/PACS\footnote{{\it Herschel Space Observatory} Photoconductor 
Array Camera and Spectrometer (PACS) \citep{pil10,pog10}.} have revealed
powerful molecular outflows in ULIRGs, traced by hydroxyl (OH), with
velocities exceeding 1000 km/s in some sources and mass 
outflow rates of at least several hundreds \Msun\ yr$^{-1}$ 
\citep[][hereafter GA14 and GA17]{fis10,stu11,gon14,gon17}. The outflows seen
in low excitation lines of OH were found to be ubiquitous in
local ULIRGs \citep[Fig.~\ref{art},][]{spo13,vei13,sto16}. These
investigations also revealed a 
correlation between the terminal outflow velocity and the 
active galactic nucleus (AGN) luminosity.
The molecular outflows are also widely observed at (sub)millimeter 
wavelengths in lines of CO, HCN, HCO$^+$, and CS 
\citep{sak09,fer10,fer15,ala11,ala15,cic12,cic14,das12,aal12,aal15,com13,lin16,gar15,per16,ima16,pri17,vei17}
and in ro-vibrational H$_2$ lines at near-IR wavelengths \citep{rup13a},
providing evidence for molecular feedback in both AGN and SB sources up to
scales of $\gtrsim1$ kpc. Recently, modeling of outflows observed in OH in
local ULIRGs has shown that their mechanical power and momentum fluxes are 
in most sources $(0.1-0.3)$\% of $L_{\mathrm{IR}}$ and 
$(2-5)\times L_{\mathrm{IR}}/c$, respectively, although some sources with AGN
luminosities at or above the quasar level significantly exceed these 
values (GA17). Depletion timescales are estimated to be $<10^8$ yr and,
notably, shorter than the consumption timescales. Comparison with CO indicates
that the OH observations are reliably calibrated with an overall abundance 
$\mathrm{[OH]/[H]}=2.5\times10^{-6}$ \footnote{This
abundance would be scaled in high-$z$ sources according to the metallicity
studies to be performed also with {\it SPICA} \citep{fer17}.}.

At high redshifts, however, direct evidence for outflows and the impact
  of feedback, particularly on the dense ISM, is much less
  constrained. Observations of outflows in high-$z$ sources have been  
widely reported in lines of ions 
\citep[e.g.][]{vil11,far12,can12,har12,bar13,gen14,carn15,zak16,nes16}, 
indicating that AGN feedback in the warm ionized phase is common at high $z$.
However, the outflowing molecular phase is basically unknown with only
a few (possible) detections reported so far
\citep[][]{pol11,nes11,gea14,falg15,feru17}, 
which limits our knowledge of a key physical process that may strongly impact
galaxy evolution on cosmic timescales -at least of luminous systems. While in
the local Universe the most powerful molecular outflows are observed in
AGN-dominated merger ULIRGs,  
analysis of deep {\it Hubble} fields based on structural parameters indicate a
merger fraction of massive ($>10^{10}$ \Msun) galaxies increasing with
redshift up to $\gtrsim10$\% at $z=1.2$, implying that a typical present-day
massive galaxy has undergone $\sim1$ major merger during the last $\sim9$ Gyr
\citep{con09,rob10,xu12}. This merging rate is capable of explaining the
build-up of the red sequence during the last $\sim8$ Gyr 
\citep{eli10,xu12,pri13},
and it is in general concordance with the $\Lambda$CDM galaxy mass assembly
model \citep{con14}\footnote{This may contradict 
  the observed fraction of present-day large spirals unless their fragile
  disks, easily destroyed in a major meger, regenerate efficiently
  \citep[e.g.][]{ham09,pue12}.}. Alternatively, smooth cold mode
accretion of intergalactic gas, which is expected from simulations to be the
main gas supply for star formation over cosmic times 
\citep[e.g.][]{mur02,ker05,ker09},  
could conceivably present accretion ``peaks'' at high redshifts generating
bright nuclear SBs and AGN activity, and thus powerful
feedback. In either case, the increasing importance of feedback processes at
high redshifts is also expected from the sharp increase of the IR luminosity
function as derived from the Spectral Energy Distribution (SED) analysis of
{\it Herschel}/PACS selected galaxies \citep{gru13}, as well as from the
increase in AGN activity \citep{hop07,mer08,del14}.
{\it SPICA} would indeed measure with unprecedented
accuracy and depth the cosmic evolution of AGN activity using both
spectroscopic and photometric techniques 
\citep[see the accompanying papers][]{spi17,gru17}.


Our very limited knowledge of the role of feedback on galaxy evolution at
high redshifts leaves crucial questions unanswered until observations of
outflow characteristics as a function of redshift can be obtained: \\
1) What is the incidence of molecular outflows during the last $\sim10$ Gyr
(up to $z\sim1.5$) from the peak of star formation and SMBH accretion
activity? How are these outflows related to the steep decline in  
star formation during this epoch? In the framework of
the two-fold evolutionary dichotomy for IR galaxies proposed by \cite{gru13},
with a strong SB-dominated phase accompanied by SMBH growth leading to
ellipticals, and a more moderate mode of star-formation that may generate
the local spiral population, are feedback processes mainly responsible for the
evolution along the SB-SMBH branch? An estimate of the outflow
luminosity/momentum function is needed to answer this question. \\ 
2) In the local Universe, massive molecular outflows in OH are observed in
(U)LIRGs, which present specific star-formation rates (sSFR) well above Main
Sequence (MS) galaxies at $z\le0.1$ \citep[e.g.][]{elb07}; but what is the
situation at $z\sim1-1.5$?
Are outflows found only in sources
above the MS, or do they also take place in MS galaxies,
i.e. in disk galaxies with “normal” rates of star formation per stellar mass?
\cite{gen14} have found that 2/3 of the most massive galaxies (both MS and
non-MS) at $z=1-3$ (44 sources) display broad ($\sim1000$ km/s) nuclear
emission in H$\alpha$, [S {\sc ii}], and [N {\sc ii}] lines suggesting 
AGN feedback, but very little is known about the molecular
component. \\
3) What is the physical link between AGN feedback on small scales and
the molecular outflows seen on large spatial scales ($>100$ pc)? Is there any
correlation between the feedback and black hole accretion and SB activities,  
{\em to be measured also in the far-IR} \citep[see][]{spi17,gru17} 
as well as in X-Rays
\citep[{\it Athena} observatory in late 2020s,][]{nan13}? 
Is the coupling due to expanding bubbles driven by ultra-fast outflows (UFOs)
that are generated in accretion disks around the SMBH, as observationally
inferred in a few sources \citep{tom15,fer15,vei17} and theoretically
well described \citep[e.g.][]{fau12,kin15,ric17}, or due to radiation
pressure on dust grains \citep{mur05,rot12,tho15,ish15}?  
What are the roles of nuclear SBs and of radio jets on molecular
outflows? \\
4) What is the physical environment (dust temperatures, continuum
optical depths at IR wavelengths, radiation densities, enclosed gas
masses, the gravitational potential well, isotopic enrichment and
metallicities, star-formation rates) in the intermediate regions surrounding
the SMBH ($\sim100$ pc)? Observations indicate that quenching of star
formation proceeds “inside-out” \citep{tac15}, so it is crucial to
observationally trace all regions of the quenching activity.  Far-IR
spectroscopy of a small sample of local galaxies with {\it Herschel}/PACS 
have shown extremely high excitation of OH, H$_2$O, and other molecular
species in most ULIRGs and in some LIRGs, seen in high-lying absorption lines
that are pumped through absorption of far-IR photons. Even though 
the regions probed by these lines are not resolved with far-IR facilities,
absorption profiles in
multiple high-excitation levels enable the estimation of the dust
temperatures in these inner regions and provide direct proof that the
circumnuclear environments in most local 
ULIRGs are warm and optically thick even at far-IR wavelengths 
\citep[$T_{\mathrm{dust}}>60$ K, GA14,][hereafter GA15]{gon15}. A source
with $R=100-170$ pc, $T_{\mathrm{dust}}=90-70$ K, and $\tau(100\mu m)=1$ emits
$10^{12}$~\Lsun, constraining the spatial scales that are probed in ULIRGs
with high-lying absorption lines in the far-IR. What are the properties of the
circumnuclear regions, locally characterized by extremely high brightness and
gas surface densities and representing the most buried stages of SB-AGN
co-evolution, in high-z ULIRGs? \\
5) How is the thermal energy and dynamical state of the molecular gas affected
by the outflow?  In particular, how is star formation in the molecular
reservoir impacted by feedback from the AGN and SB? How is
this molecular reservoir chemically affected when loaded by the outflow?
Specifically, what is the ionization rate of the molecular gas?

\begin{figure}
\begin{center}
\includegraphics[width=8cm]{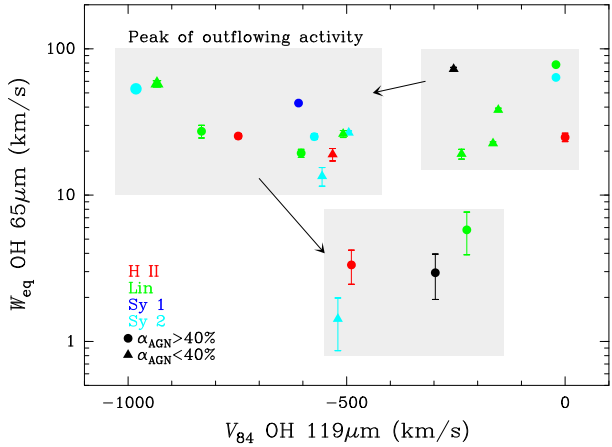}
\caption{The equivalent width of the OH65 doublet at systemic velocities
  ($-200$ to $+200$ \kms), probing buried and warm sources in the far-IR, as a
  function of $V_{84}$ (OH119) --$84$\% of the absorption in the OH119 doublet
  is produced at velocities more positive than $V_{84}$, so that this quantity
  is a measure of the outflow velocity \citep{vei13}. A possible
  evolutionary sequence is depicted, in which the peak of outflowing activity,
  characterized by high outflowing velocities within a (still) buried phase,
  is preceded by an extremely buried phase (a ``greenhouse'' galaxy) with low
  velocities (and dominated by accretion in some cases) and followed by a
  stage where the nuclear columns of gas have decreased and the outflowing
  activity has subsided (adapted from GA17). 
} 
\label{oh65_local}
\end{center}
\end{figure}


\section{The promise of {\it SPICA}}

With its expected exceptional sensitivity and spectroscopic capabilities 
\citep{swi09,nak14,sib16,roe17,spi17}, the far-IR observatory {\it SPICA} 
would address these questions by spectroscopically observing galaxies from
$z=0$ to $z\sim1.5$ in multiple molecular and atomic lines. 
The SpicA FAR Infrared instrument (SAFARI) \citep{swi09,pas16,spi17},
covering the wavelength range $34-230$ $\mu$m, is designed to provide two
spectroscopic observing modes, the low-resolution (LR) mode with nominal 
$R_{\mathrm{nom}}=300$, and the high-resolution (HR) mode with 
$R_{\mathrm{nom}}=2\times10^3\times200\mu m/\lambda$.
As shown below, a ULIRG with $L_{\mathrm{IR}}\sim3\times10^{12}$
L$_{\mathrm{\odot}}$ at $z=1.5$ would only require 
$\sim2$ h of observing time in the LR mode to 
obtain 1$\sigma$ noise of $\sim1$\% the continuum level, enabling the
detection of molecular outflows (as seen in the local Universe)
including the simultaneous detection of lines of OH and H$_2$O, and thus
enabling the exploration of multiple transitions of several species in
significant samples of galaxies.  
From multitransition observations of OH, H$_2$O, OH$^+$, and other molecules
(mostly light hydrides), the nuclear molecular outflow and related
characteristics, e.g. the properties of the far-IR continuum emission, 
the column densities, mass outflow rates, momentum boost, depletion
timescales, and the molecular richness would be probed by {\it SPICA} as a
function of redshift.   

Ground-based interferometric observations at (sub)millimeter wavelengths have
also traced molecular outflows in local ULIRGs with CO emission line wings
\citep[e.g.][]{fer10,cic12,cic14,gar15} and HCN-HCO$^+$ wings
\citep[e.g.][]{sak09,aal12,aal15,lin16,pri17}. In 
the next years, ALMA and NOEMA in the (sub)millimeter, and the X-ray
observatory {\it Athena} in late 2020s, will continue observing
outflows in galaxies. Nevertheless, the observations in the far-IR domain by
{\it SPICA} would be crucial and complementary, as they would provide unique
information that cannot be obtained at other wavelengths:

$(i)$ Far-IR observations unambiguously identify the outflow kinematics
through P-Cygni profiles or blueshifted absorption wings, ruling out 
other possibilities as inflow, injection of turbulent energy \citep{gui15},
or motions associated with merging.
Because the blueshifted line wings are observed in absorption
in the far-IR, low-velocity ($100-300$ km/s) outflowing gas that 
carries a significant fraction of the outflowing mass in ULIRGs, would also
be identified and analyzed with a sensitive far-IR observatory with 
high spectral resolution. This
component may be missed with pure emission lines because of confusion
with the line core. While molecular P-Cygni and absorption line shapes are
also observed in the submillimeter toward extremely buried galaxy cores
\citep{sak09,ima16}, the far-IR domain provides a very sensitive probe of
outflows owing to both the increase of continuum optical depth with decreasing
wavelengths and to the high transition probabilities of light hydrides. 
The far-IR observations also probe the physical and chemical conditions in the
circumnuclear regions on a few $\times100$ pc scales through the observation
of high-lying lines in absorption at systemic velocities 
\citep[][GA14, GA15, GA17]{gon12,fal15,fal17},
thus enabling the characterization of the spatial components of the observed
SED at the wavelengths where most of the galaxy luminosity is emitted. This
connection between lines and SEDs is an exclusive virtue of far-IR
spectroscopy.

$(ii)$ While {\it Athena} will probe the extremely
hot phase of the outflows around the SMBHs, and ALMA will explore the cold
phase (with energy levels at typically up to few tens K), the far-IR
observatory {\it SPICA} would probe the intermediate molecular/warm phase and
hence the key link between the innermost (pc-scale) and the outermost
(kpc-scale) regions. 
In its most extended configuration ($\sim16$ km), ALMA provides an
angular resolution of $\sim15$ mas at 300 GHz ($\sim100$ pc at $z=0.5$). By
measuring multiple energy level line profiles with sufficient signal-to-noise,
{\it SPICA}/SAFARI -despite its low angular resolution, $\sim4.5''$ to $19''$-
would probe similar spatial scales for the most excited outflowing components.
This is because the high-lying lines (pumped through the far-IR field) require
very high far-IR radiation densities, and high \tdust$-\tau(100\mu m$) implies
small sizes for given luminosities  (GA14 \& GA17 and references therein).

$(iii)$ The circumnuclear environments that are traced in the far-IR provide
relevant clues to interpret global galaxy properties such as the 
line deficits, the far-IR colors in galaxies, as well as their layered
structure (i.e. temperature and column density as a function of extent;
GA12). At $z<1$, a very rich, unique chemistry can be studied in the far-IR 
through the observation of high-lying lines of OH, H$_2$O, CO, OH$^+$,
H$_2$O$^+$, H$_3$O$^+$, NH, NH$_2$, NH$_3$, CH, CH$^+$, C$_3$, etc. The
excited lines of the O-bearing molecular ions are specially interesting as
they directly probe the ionization rate of the gas due to cosmic rays and/or
X-rays in the nuclear region \citep{gon13}, filtering out more
spatially extended environments that are best probed with the ground-state
lines \citep{tak16}. The best tracers of photon dominated
regions (PDRs) and the corresponding kinematics, i.e. [O {\sc i}]63-145$\mu$m
and [C {\sc ii}]157$\mu$m, also lie in the far-IR, and outflowing gas has been
detected in the [C {\sc ii}]157$\mu$m line in local ULIRGs \citep{jan16} and
in a bright quasar at $z > 6$ \citep{mai12,cic15}.


\begin{figure*}
\begin{center}
\includegraphics[width=17cm]{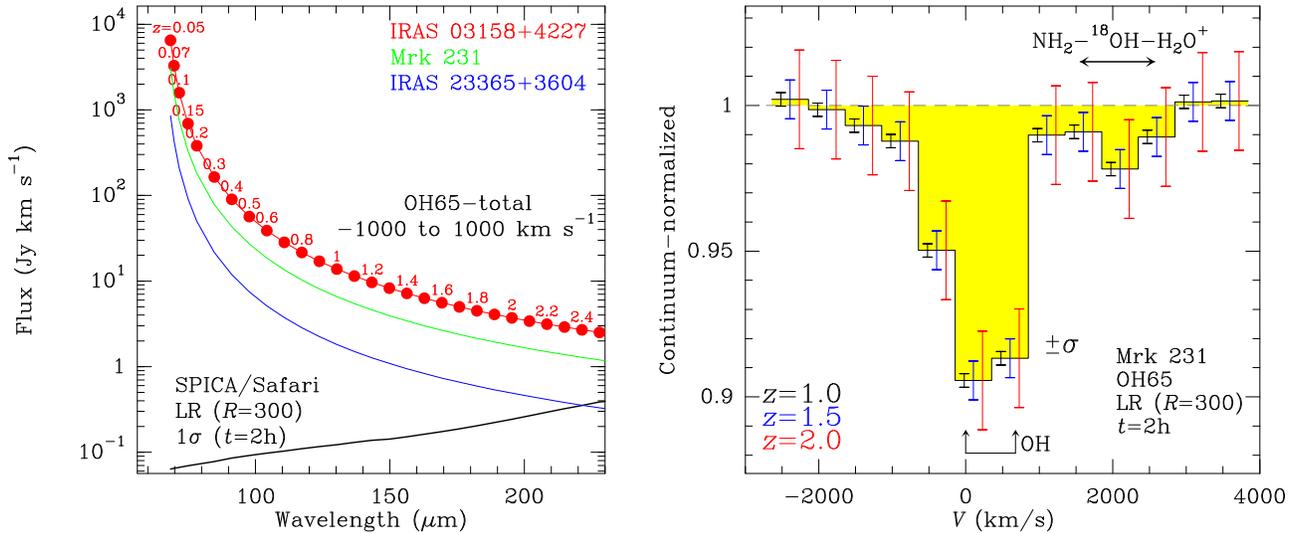}
\caption{
{\it Left}: Predicted integrated absorbing flux of the OH65 doublet in three
local ULIRGs (IRAS~03158+4227, Mrk~231, and IRAS~23365+3604, from $-1000$
\kms\ to 1000 \kms) as a function of redshift (red numbers) and observed
wavelength (abscissa). The black curve shows the expected sensitivity 
($1\sigma$) of {\it SPICA}/SAFARI LR ($R=300$) for 2 hours of observing time,
indicating that the doublet, probing buried stages, would be easily detected
in similar ULIRGs up to $z=1.9-2.5$. {\it Right}: The continuum-normalized
OH65 spectrum of Mrk~231 observed with {\it Herschel}/PACS smoothed to the
resolution of {\it SPICA}/SAFARI LR, with 2 spectral points per resolution
element. The three errorbars in each spectral channel indicate the $\pm\sigma$
uncertainty expected with SAFARI for 2 hours of observing time at the selected
redshifts of $z=1.0$, $1.5$, and $2.0$. The weak absortion around
$V\sim2000$ \kms\ is a blend of NH$_2$, $^{18}$OH, and H$_2$O$^+$ lines.} 
\label{oh65_lr}
\end{center}
\end{figure*}

Lastly and importantly, far-IR spectroscopy in these same transitions opens
the exciting possibility of the detection of the feeding of the galaxy cores:
at low redshifts, inflowing gas will be identified through
inverse P-Cygni profiles or redshifted absorption wings of OH and 
[O {\sc i}]63$\mu$m, as seen in the local NGC 4418, Zw 049, Arp~299a,
IRAS~11506$-$3851, IRAS~15250+3609, and Circinus 
\citep[][GA17]{gon12,fal15,fal17,sto16}.
While inflowing motions can also be inferred from skewed profiles of pure
emission lines and HI redshifted absorption \citep{cos13}, the far-IR
offers a very sensitive probe of inflow due to the increasing
continuum optical depth. 

In summary, while extraordinary results will be obtained in the years to come
from observatories working on a wide range of wavelengths, 
there is no replacement for the unique capabilities of 
the far-IR counterpart {\it SPICA} in the late 2020s. 

\begin{figure*}
\begin{center}
\includegraphics[width=17cm]{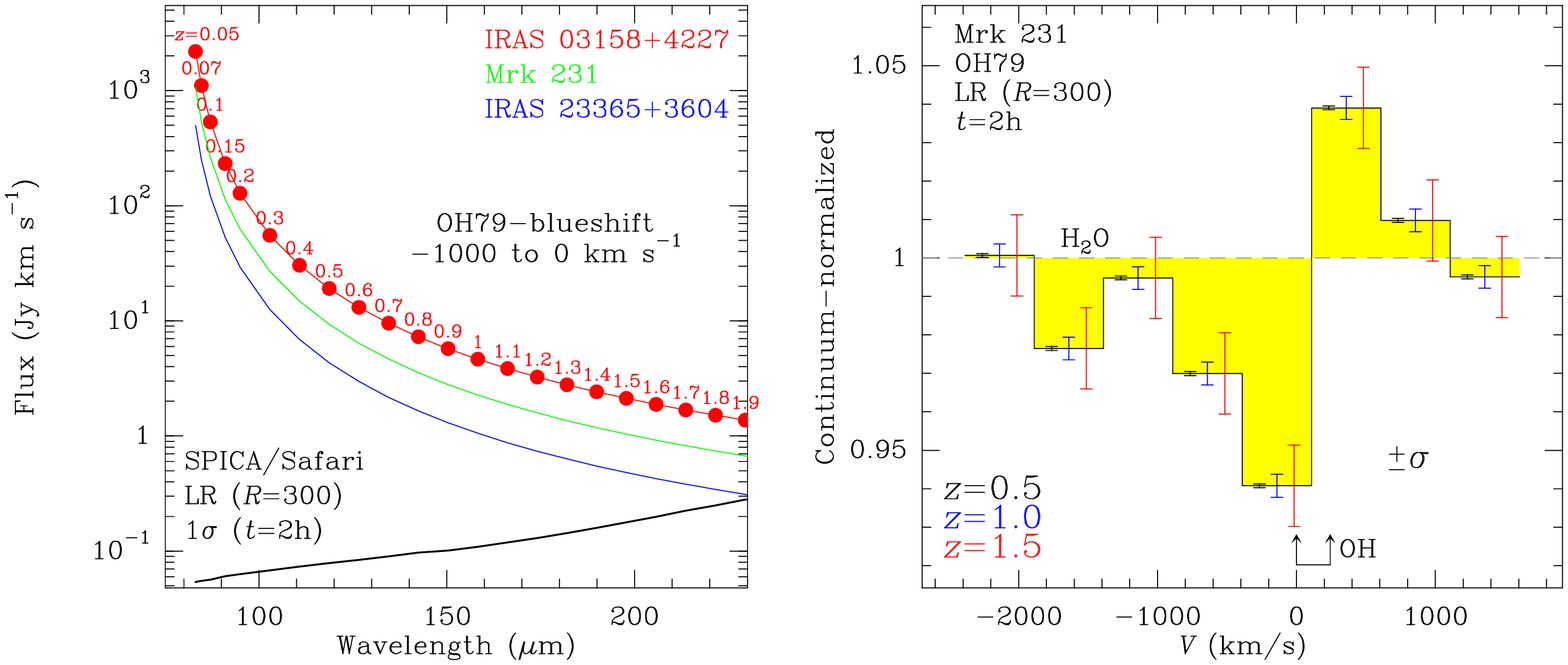}
\includegraphics[width=17.2cm]{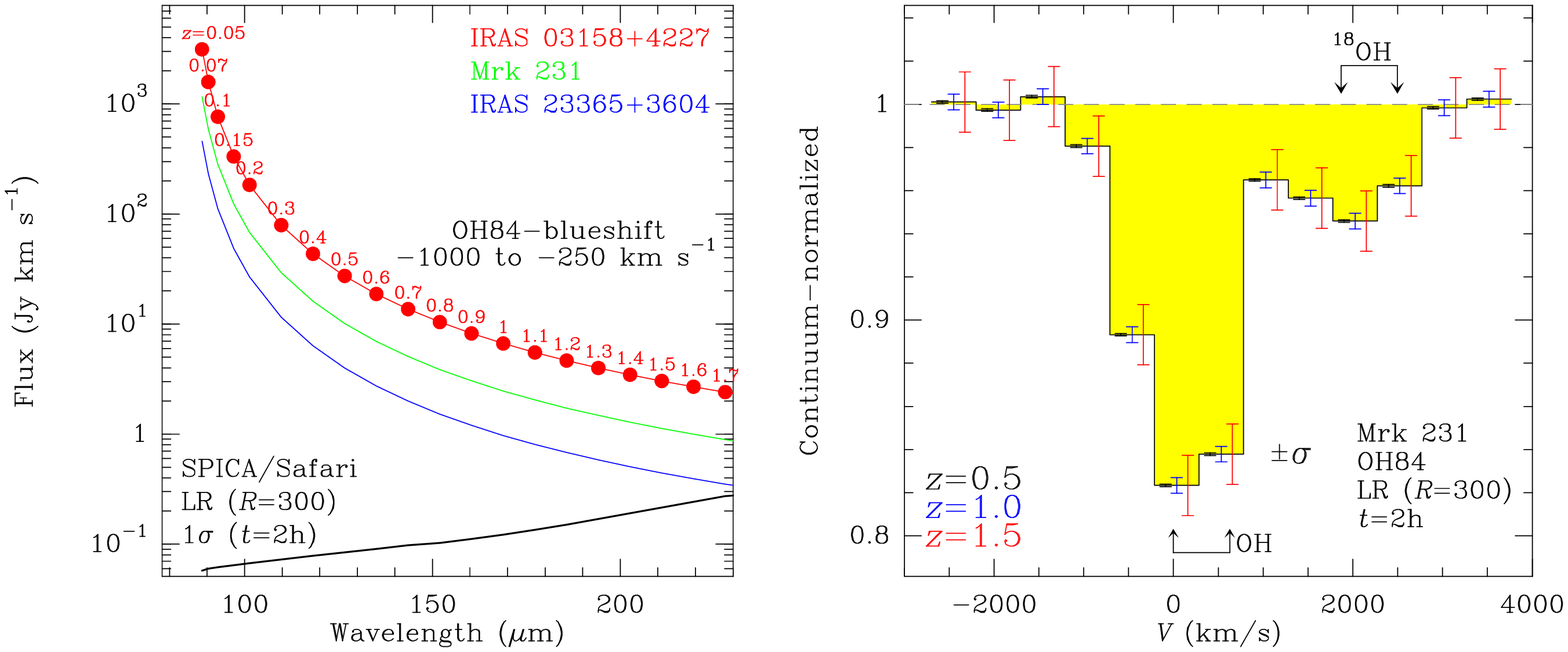}
\caption{{\bf Upper panels.}
{\it Left}: Predicted integrated absorbing flux of the OH79 doublet in
three local ULIRGs (IRAS~03158+4227, Mrk~231, and IRAS~23365+3604, all showing
P-Cygni profiles in OH79) at blueshifted velocities 
(from $-1000$ \kms\ to 0 \kms) as
a function of redshift (red numbers) and observed wavelength (abscissa). 
The black curve shows the sensitivity ($1\sigma$) expected for 
{\it SPICA}/SAFARI LR ($R=300$) with 2 hours of observing time, indicating
that molecular outflows would be easily detected in ULIRGs up to $z=1.3-1.9$. 
{\it Right}: The continuum-normalized OH79 spectrum of Mrk~231 as observed
with {\it Herschel}/PACS smoothed to the
resolution of {\it SPICA}/SAFARI LR, with 2 spectral points per resolution
element. The three errorbars in each spectral channel indicate the $\pm\sigma$
uncertainty for SAFARI with 2 hours of observing time at the selected
redshifts of $z=0.5$, $1.0$, and $1.5$. The absortion at $V<-1300$ \kms\ is
due to \hdo\ $4_{23}-3_{12}$. Note that not only the blueshifted absorption 
wing would be detected, but also the redshifted emission feature
(i.e. P-Cygni), unambiguously revealing outflowing gas.
{\bf Lower panels.} The corresponding predictions for the excited OH84
doublet. Fluxes are shown for velocities between $-1000$ and $-250$ \kms. In
the right-hand panel the absorption around $2000$ \kms\ is due to $^{18}$OH
with possible contribution by NH$_3$ in some sources (see GA12).
} 
\label{oh79_lr}
\end{center}
\end{figure*}

\begin{figure*}
\begin{center}
\includegraphics[width=17cm]{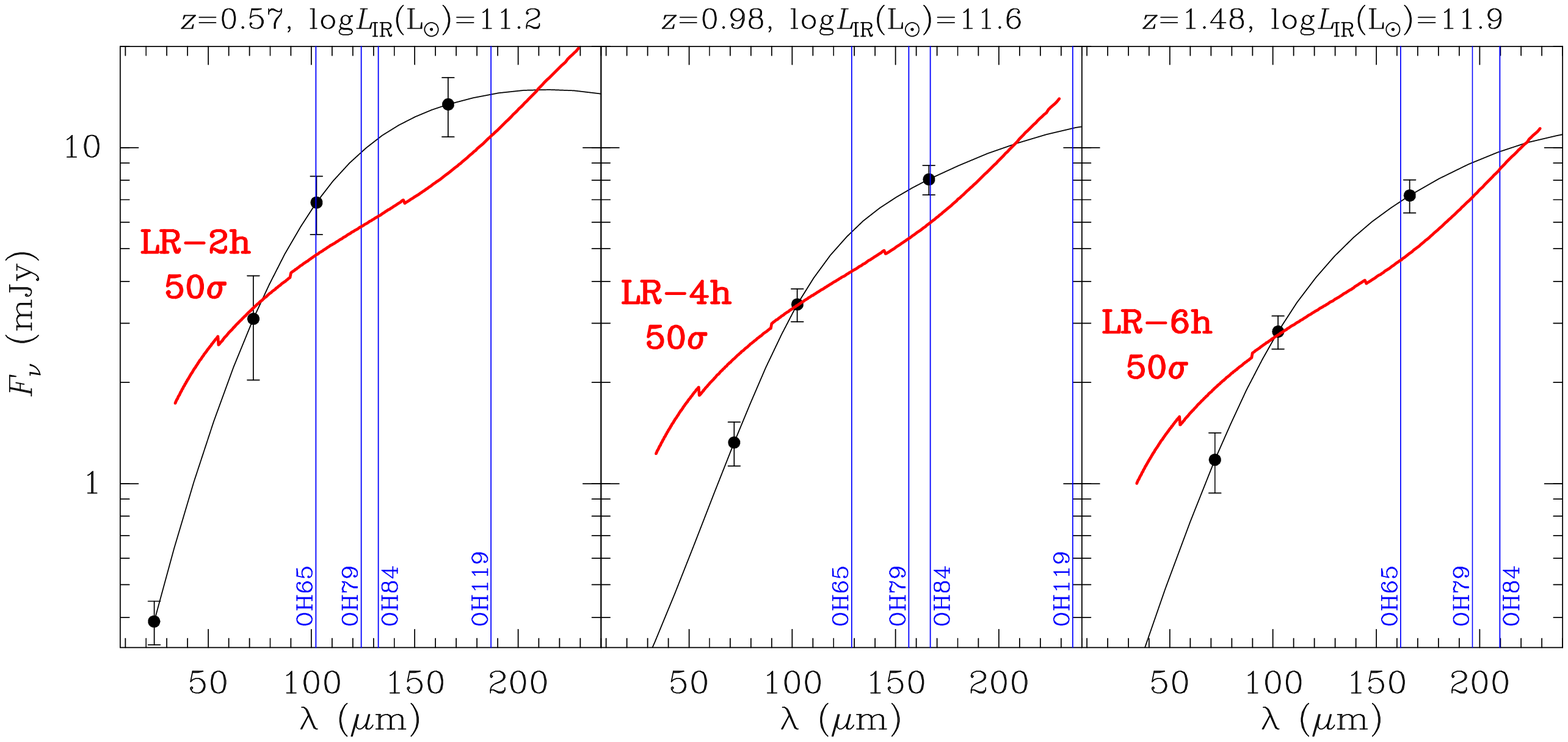}
\caption{The black points with errorbars and black curves show the SED of Main
  Sequence galaxies in the highest $\mathrm{log}\,M_*(\mathrm{M_{\odot}})=11.2$
  bin at different redshifts \citep[from][]{sch15}, and the vertical
  blue lines indicate the observed wavelengths of the OH65, OH79, OH84, and
  OH119 doublets. The red curves show the expected $50\sigma$ sensitivities in
  LR mode and for one spectral channel attained in $2-6$h, indicating the
  capability of the designed {\it SPICA}/SAFARI instrument to explore the
  possible outflow origin of the ``bending'' of the 
  $M_*$-SFR (MS) correlation in the high $M_*$ bin.
} 
\label{ms}
\end{center}
\end{figure*}

\subsection{Molecular outflows in the far-IR: P-Cygni profiles and blueshifted
  absorption wings}

We adopt 3 local template ULIRGs to make predictions on the
detectability of molecular outflows at high-$z$, by 
scaling the far-IR spectroscopic observations obtained with 
{\it Herschel}/PACS: IRAS~03158+4227, Mrk~231, and IRAS~23365+3604 (GA17). 
The IR luminosities ($8-1000$ $\mu$m, $L_{\mathrm{IR}}$) are listed in
Table~\ref{tab1}. All three sources, as well as many other local ULIRGs, show
clear evidence for massive molecular outflows from P-Cygni line profiles in
the ground-state OH119 and OH79 doublets \citep[e.g.][]{fis10}. In the cases
of IRAS~03158+4227 and Mrk~231, high-velocity ($\gtrsim1000$ \kms) absorption
wings are also detected 
in the excited OH84 and OH65 doublets, indicating compact outflowing gas with
high column densities (GA17). We use the currently expected sensitivities of
the {\it SPICA}/SAFARI instrument as shown in Appendix~\ref{appa} and 
described in \cite{roe17}.

\begin{table}
\caption{Local template ULIRGs used to make {\it SPICA} detectability
  predictions at high redshift.} 
\begin{center}
\begin{tabular}{lc}
\hline\hline
Galaxy & $L_{\mathrm{IR}}$ \\
       & ($10^{12}$ L$_{\mathrm{\odot}}$) \\
\hline
 IRAS~03158+4227  & $4.3$ \\
 Mrk~231          & $3.4$ \\ 
 IRAS~23365+3604  & $1.4$ \\ 
\hline\hline
\end{tabular}
\end{center}
\label{tab1}
\end{table}

\subsubsection{Identifying massive molecular outflow candidates}
\label{sec:buried}

It has been proposed that the most active stages of SMBH growth and nuclear SB
co-evolution, involving high columns of 
gas funneled towards the nuclear regions of galaxies mostly as a consequence
of merging or interactions, are also the stages in which feedback is
expected to be most powerful, regulating the SMBH growth and star-formation
burst. A unique, unambiguous probe of this buried stage comes from the
absorption at far-IR wavelengths in high-lying transitions of hydrides,
requiring both high dust temperatures and molecular columns over relatively
compact regions (less than a few $\times100$ pc). The OH65 doublet, with 
$E_{\mathrm{low}}\approx300$ K, is particularly well suited for the
identification of these 
buried and warm sources, because it can only be excited through absorption of
far-IR photons in these type of environments and is correlated in local
galaxies with global galaxy properties: the [C {\sc ii}]158$\mu$m deficit, the
luminosity-to-gas mass ratio, the silicate absorption feature at 10 $\mu$m,
and the 60-to-100 $\mu$m far-IR color (GA15). In addition, we show in
Fig.~\ref{oh65_local} the relationship found in local sources between the
equivalent width of the OH65 doublet at systemic velocities and the velocity
of the outflowing molecular gas as traced by the OH119 doublet (GA17),
indicating that indeed the highest (more blueshifted) observed outflowing
velocities are found in sources with high OH65 absorption
($W_{\mathrm{eq}}>10$ \kms). This is entirely consistent with the finding that
the highest outflow velocities in OH are found in sources with strong silicate
absorption \citep[the upper branch of the fork diagram in][]{spo07,spo13}.

The OH65 doublet provides a unique tool to easily identify high nuclear
activity, that is, the most compact and buried sources which pinpoint a
  particular phase in the evolution of galaxies where feedback might be
  expected to be strong. 
While {\it Herschel}/PACS has observed this doublet in only $\sim30$ local
galaxies, Fig.~\ref{oh65_lr} shows that the population of ULIRGs capable
  of generating strong absorption in OH65 would be identified up to 
$z\sim2$, i.e. out to $3/4$ of the Hubble time, through observations
  with {\it SPICA}/SAFARI in LR mode with only 2 hours of observing time per
  source. Details on flux calculations based on currently expected
SAFARI sensitivities are given in Appendix~\ref{appa}.  


\subsubsection{Molecular outflows in OH at $z>1$: the LR mode}
\label{sec:highz}

The extremely efficient {\it SPICA}/SAFARI LR mode would enable the detection
of molecular outflows at moderately high-$z$. The ground-state OH79
doublet, showing P-Cygni line shapes and correlated with the CO 1-0
luminosities in the blueshifted line wings in local ULIRGs
\citep[][GA17]{stu11}, would be observable with {\it SPICA}/SAFARI up to
$z\approx1.9$ and provides the best tool for identifying molecular feedback. 
We compare in Fig.~\ref{oh79_lr}
(upper-left panel) the predictions for the blueshifted OH79 absorption flux as
a function of $z$ with the expected {\it SPICA}/SAFARI sensitivity in LR, with
only 2 hours of observing time (see details in Appendix~\ref{appa}). Strong
outflow sources as IRAS~03158+4227 and Mrk~231 would be detected up to the
maximum observable redshift (set by the instrumental limit), and more
moderate outflow sources like 
IRAS~23365+3604 would be detectable up to at least $z\approx1.3$. The
upper-right panel of Fig.~\ref{oh79_lr} shows that 
{\em the LR mode predicts a clear P-Cygni line
  shape for the OH79 doublet in outflowing sources, thus unambiguously
  revealing massive molecular outflows}\footnote{In sources like
  IRAS~23365+3604, signal will be obtained in two consecutive channels in
  absorption and emission, and outflow detection at $4\sigma$ level will rely
  on $\sim3\sigma$ detection in both channels.}.

{\em Excited} outflowing gas would also be detected through the OH84 doublet
(lower panels in Fig.~\ref{oh79_lr}). In the LR mode, the two $l-$doubling
components are blended in one single feature, which shows a ``blue'' 
asymmetry characteristic of outflowing gas with predicted fluxes well above
the sensitivity limits up to at least $z\approx1.3$. This asymmetry is also
apparent in the OH65 doublet (Fig.~\ref{oh65_lr}).

Observations of the cross-ladder, ground-state OH doublets at 35 and 53.3
$\mu$m would enable the exploration of sources at $z>2$. The equivalent width
of the OH35 doublet in Mrk~231, detected with {\it Spitzer}/IRS, is
$\approx50$ \kms, and would be detected at $z=3$ with 5$\sigma$ confidence
using the LR mode and 10 hours of observing time. The observation would detect
simultaneously the OH53.3 at 8$\sigma$ level, though the blueshifted wing in
this doublet is blended with highly excited lines of \hdo\ and OH (GA14).

We also remark that hyper-luminous galaxies like HFLS3 
\citep[$L_{\mathrm{FIR}}\sim3\times10^{13}$ \Lsun][]{rie13}
would be easily detected with {\it SPICA}/SAFARI in the LR mode. This
extraordinary source at $z=6.3$, rich in molecular lines including OH, H$_2$O,
and OH$^+$, shows a continuum flux density of 20 mJy at
$\lambda_{\mathrm{rest}}\sim150$ $\mu$m, expectedly similar to the flux
density at 35 $\mu$m. In such luminous galaxies with broad molecular
linewidths ($500-1500$ \kms\ except CO 1-0), the OH35 doublet would be
detected in LR mode up to the maximum observable redshift,
$z\approx5.5$, possibly showing blueshifted line wings in absorption. 
Some \hdo\ lines at shorter wavelengths could be also detected at even higher
$z$. Likewise, gravitational lensing enables the detection of
submillimeter lines of H$_2$O, H$_2$O$^+$, and CH$^+$ in $z=2-4$ galaxies
\citep{omo11,omo13,yan13,yan16,falg15} and will enable the 
studies of the molecular phase in splendid detail with the 
{\it SPICA} observatory.

\begin{figure*}
\begin{center}
\includegraphics[width=17cm]{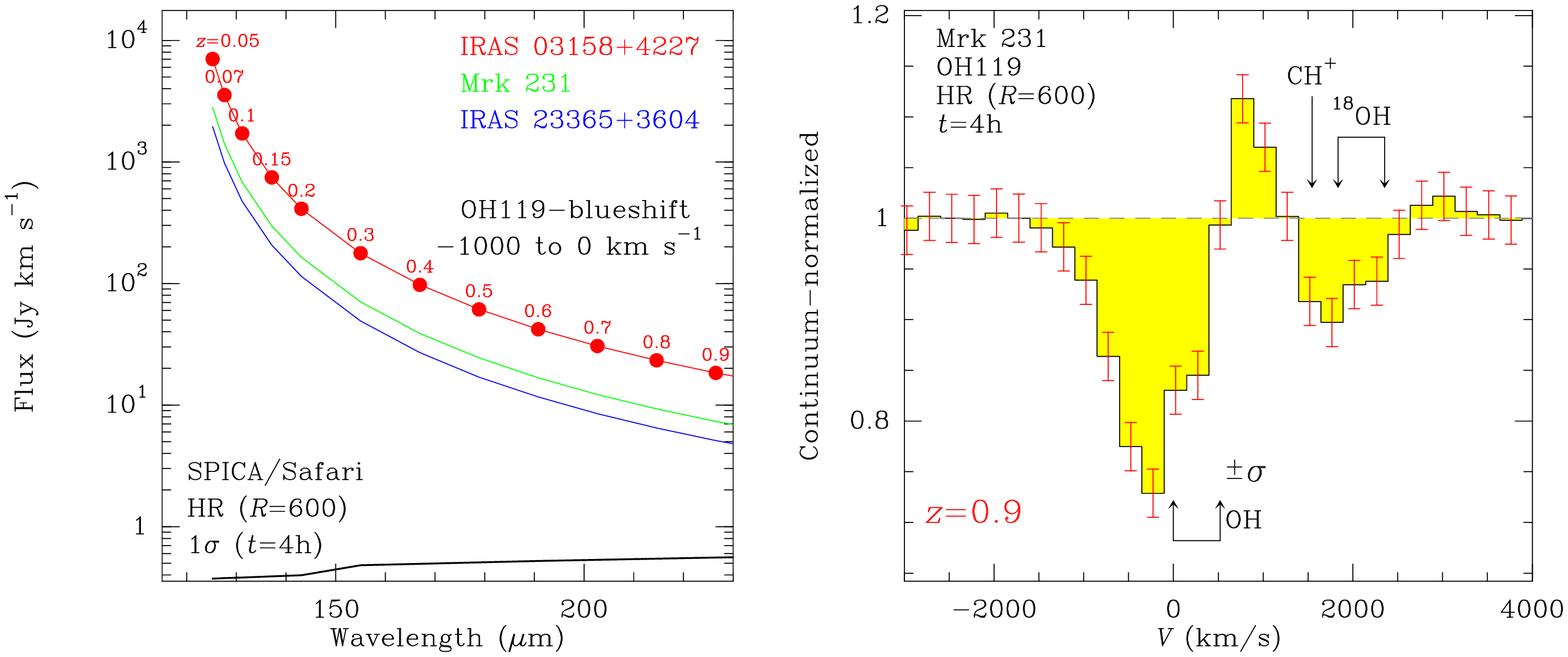}
\includegraphics[width=17cm]{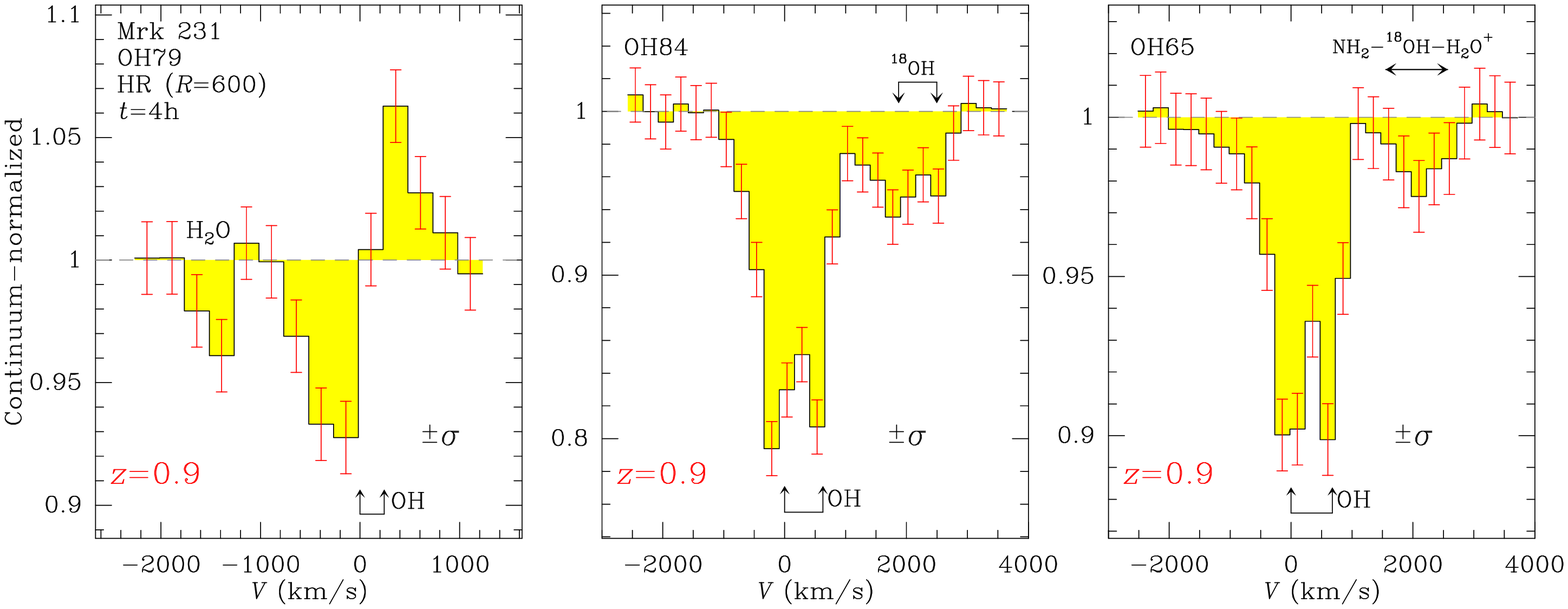}
\caption{{\bf Upper panels.}
Same as Fig.~\ref{oh79_lr} for the OH119 doublet in HR mode smoothed to a
spectral resolution of $R=600$ and with 4 hours of observing time,
illustrating the high-quality spectra that would be obtained 
with {\it SPICA}/SAFARI in this OH
doublet up to the maximum observable redshift, $z\approx0.94$.
Contribution to the absorption by $^{18}$OH would also be detectable,
constraining the metallicity of the sources \citep[see also the companion
paper][]{fer17}.
{\bf Lower panels.} Continuum-normalized spectra of the OH79, OH84, and OH65
doublets in Mrk~231 as observed with {\it Herschel}/PACS with the
resolution of {\it SPICA}/SAFARI HR smoothed to $R=600$. The errorbars
indicate the expected $\pm\sigma$ uncertainty reachable with SAFARI with 4
hours of observing time at $z=0.9$.
} 
\label{oh119_hr}
\end{center}
\end{figure*}

\begin{figure*}
\begin{center}
\includegraphics[width=17cm]{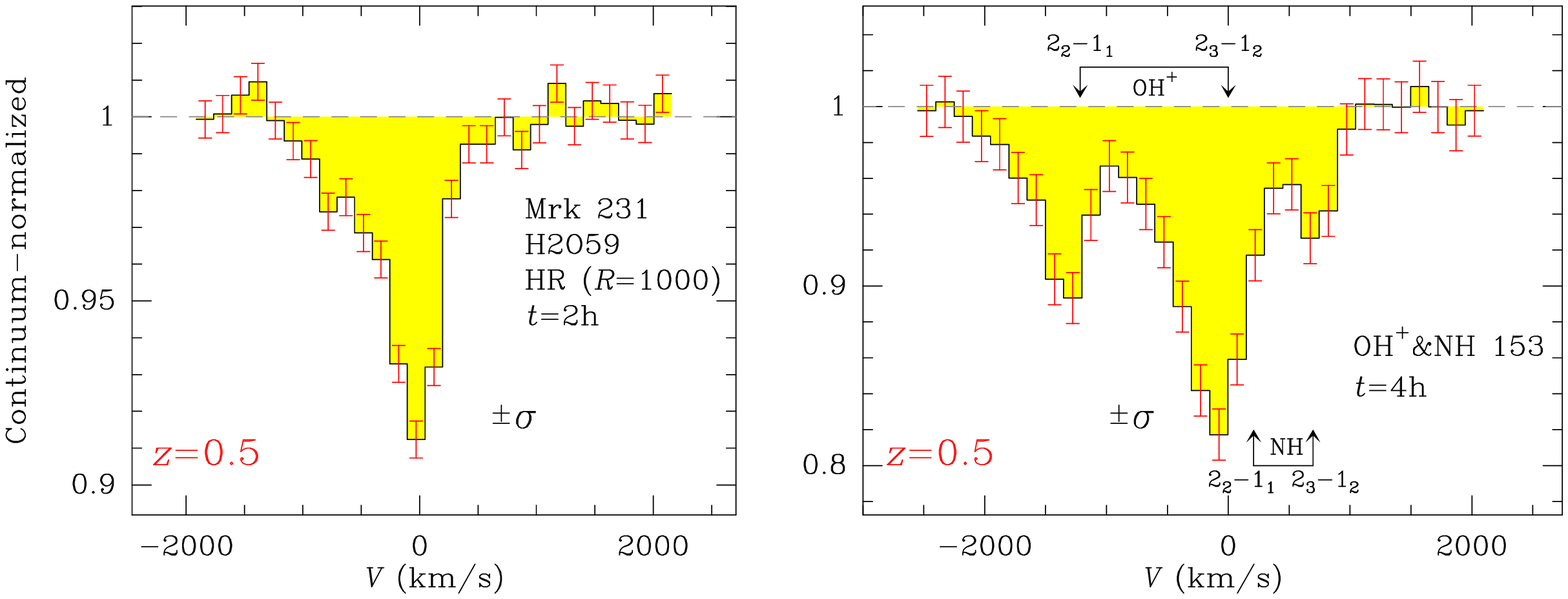}
\caption{The H$_2$O $4_{32}-3_{21}$ line at 59 $\mu$m, and the spectrum around
  153 $\mu$m including $2_{J}-1_{J'}$ lines of OH$^+$ and NH, observed with
  {\it Herschel}/PACS in Mrk~231, smoothed to a resolution of 1000 (adapted
  from Fischer et al., in preparation). The reference velocity corresponds
  the $2_{3}-1_{2}$ line of OH$^+$, which dominates over the partially blended
  NH $2_{2}-1_{1}$ line. Errorbars indicate the expected $\pm\sigma$ 
  uncertainties for {\it SPICA}/SAFARI in HR mode with 2 (left) and 4 (right)
  hours of observing time. The H$_2$O and OH$^+$ lines show blueshifted
  absorption wings indicative of outflowing gas.
} 
\label{h2ooh+}
\end{center}
\end{figure*}

\subsubsection{The bending of the Main Sequence at high $M_*$}
\label{sec:ms}

Main Sequence (MS) galaxies, dominating the IR luminosity function at all
redshifts \citep{gru13}, show the well-known correlation between the
stellar mass ($M_*$) and the star formation rate (SFR) 
from the local Universe up to at least $z\sim3$
\citep[e.g.][]{noe07,elb07,rod11,sch15}, indicating that 
most galaxies produce stars at a relatively constant rate over their
  lifetime, 
gradually declining in intensity over a significant fraction of cosmic time
\citep[e.g.][]{noe07,pen15}. 
The correlation is consistent with having a linear slope (in the log) at
  low $M_*$, then flattening at high $M_*$.  This "bending" is more pronounced
  at low redshifts \citep[e.g.][]{lee15,sch16}.
The cause of this bending has not been established, but it has been
  suggested that it might be associated with the formation of ellipticals
  through environmental quenching \citep[e.g.][]{dek06}, strangulation
  \citep{pen15}, or outflows \citep[negative feedback; e.g.][]{spr05}.

With the {\it SPICA} observatory, we would be able to address the
importance of outflows by 
taking IR spectra of a large sample of massive, MS galaxies.
The SEDs of MS galaxies 
with $\mathrm{log}\,M_*(\mathrm{M_{\odot}})=11.2$ in 3 redshift bins are shown
in Fig.~\ref{ms}, and the positions of the OH doublets at 65, 79, 84, and 119
$\mu$m (GA14, GA17) are shown with vertical
lines. With the currently expected sensitivities of the LR mode, we have shown
in Figs.~\ref{oh65_lr} and \ref{oh79_lr} that a $1\sigma$ uncertainty of
$\approx2$\% of the continuum would enable the detection of massive molecular
outflows. We thus show in Fig.~\ref{ms} that 
this level of sensitivity is reachable for high-mass MS galaxies with 
$<6$h of observing time per source. 
{\em SPICA/SAFARI would thus enable a systematic search for
  outflows in high-mass MS galaxies}.\footnote{Judging from the low 
  $f60/f100\sim0.5$ ratios in Fig.~\ref{ms}, as compared with local galaxies
  with high nuclear activity (GA15), we would not expect massive feedback in
  MS sources unless it occurred in an earlier phase. It would therefore be
  most compelling an exploration based on the OH119 doublet, i.e. up to
  $z\sim1$.} 

\subsubsection{Molecular outflows in OH at $z<1$: the HR mode}
\label{sec:lowz}

The ground-state OH119 doublet shows ubiquitous P-Cygni line shapes in
local ULIRGs \citep{spo13,vei13}, and would be detectable with 
{\it SPICA}/SAFARI ($\lambda_{\mathrm{max}}=230$ $\mu$m) up to $z=0.94$. 
The doublet is usually strong with blueshifted absorption deeper than 20\% 
of the continuum, which would enable its observation with {\it SPICA}/SAFARI
in HR mode. Any P-Cygni line shape 
would be easily detectable with 4h of observing time, providing high-quality
spectra in this doublet (see Fig.~\ref{oh119_hr}). With $R=600$, the four
doublets OH119, OH79, OH84, and OH65 (see lower panels of
Fig.~\ref{oh119_hr}) would be analyzed enabling an estimate of the energetics 
associated with the outflows (GA17).

We remark that the strongest drop in IR luminosity density occurs at
$0<z\le1$ \citep{gru13}, and {\it SPICA}/SAFARI would potentially 
enable the determination of molecular outflow statistics during this crucial
epoch (the last $\sim8$ Gyr of the Universe) through surveys. Hundreds of
galaxies would be potentially detected in the OH119 doublet and other doublets
with $\sim500$h of observing time, enabling to constrain the outflow
luminosity function in several redshift bins up to $z\sim1$.  
In addition, the statistics of the outflow velocity
\citep{stu11,spo13,vei13,sto16}, as well as of the mass  
outflow rates, momentum and energy fluxes (GA17) as a function of
the AGN and SB luminosities would be established 
for large numbers of galaxies over more than half the age of the
  Universe.

\subsubsection{Probing the outflows with H$_2$O and their ionization rates 
with OH$^+$} 
\label{sec:h2ooh+}

H$_2$O couples very well to the IR radiation field and presents dozens of
absorption lines in the far-IR spectra of buried galaxies (e.g. GA12). 
At high redshifts, the submillimeter emission lines of H$_2$O have been
detected in several sources 
\citep{wer11,omo11,omo13,yan13,yan16,rie13,gul16}.
Based on our Mrk~231 template and the expected SAFARI sensitivities, several
H$_2$O lines with $\lambda_{\mathrm{rest}}<77$ $\mu$m would be detected up to
$z\sim2$ with the LR mode of {\it SPICA}/SAFARI in only 2 hours of observing
time. In Mrk~231, these lines peak at systemic velocities and show a
blueshifted absorption wing, which would be too weak even for {\it SPICA}  to
detect at high redshifts. Nevertheless, averaging the profiles of several
H$_2$O lines, or the profile of a given line in a sample of observed
  galaxies, could enable the detection of a blue line-shape asymmetry in 
strong sources at $z>1$. At $z<1$, the wings of the individual lines are
detectable in the HR mode with $R=600-1000$, as illustrated in
Fig.~\ref{h2ooh+} for the H$_2$O $4_{32}-3_{21}$ transition at
$\lambda_{\mathrm{rest}}=59$ $\mu$m ($E_{\mathrm{lower}}\approx300$ K).  

Figure~\ref{h2ooh+} also shows the {\it Herschel}/PACS spectrum of Mrk~231
smoothed to $R=1000$ around $\lambda_{\mathrm{rest}}=153$ $\mu$m (adapted from
Fischer et al., in preparation), showing strong absorption in two (of the
allowed six) fine-structure $2_{J}-1_{J'}$ lines of OH$^+$ and NH. The OH$^+$
lines show blueshifted absorption wings extended to $\sim-1000$ \kms\ in both
transitions, as well as in other fine-structure lines. Together with the
$3_{J}-2_{J'}$ (at $\sim100$ $\mu$m) and $4_{J}-3_{J'}$ (at $\sim76$ $\mu$m)
lines, and in combination  with lines of H$_2$O$^+$, H$_3$O$^+$, and OH, the
OH$^+$ lines {\it SPICA}/SAFARI would provide an excellent data set for
modeling the ionization rates of the molecular outflows 
\citep{gon13,tak16}.

\begin{figure}
\begin{center}
\includegraphics[width=8.2cm]{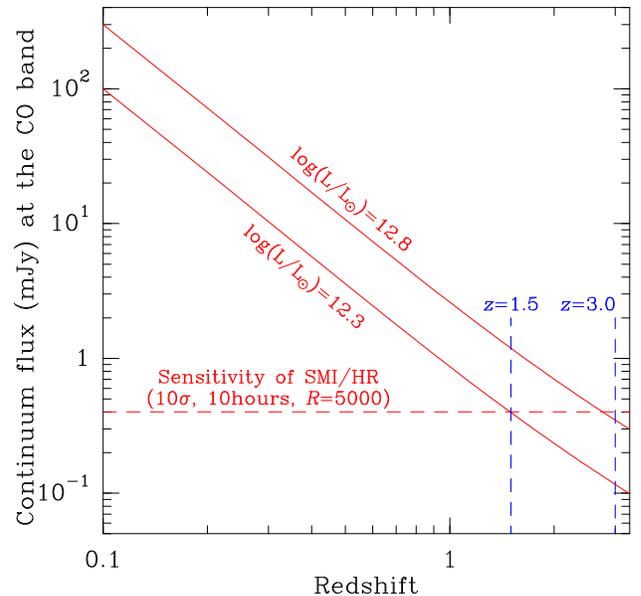}
\caption{The flux densities of ULIRGs at the wavelength of the CO band
  ($\sim4.7$ $\mu$m) together with the expected sensitivity of 
{\it SPICA}/SMI/HR with the spectral resolution binned optimal for this
study. With the spectral coverage of $12-18$ $\mu$m, which corresponds to a
redshift of $z = 1.5-3$ for the CO band, {\it SPICA}/SMI can observe this
feature toward luminous ULIRGs. 
} 
\label{coband}
\end{center}
\end{figure}

\begin{figure}
\begin{center}
\includegraphics[width=19.99cm]{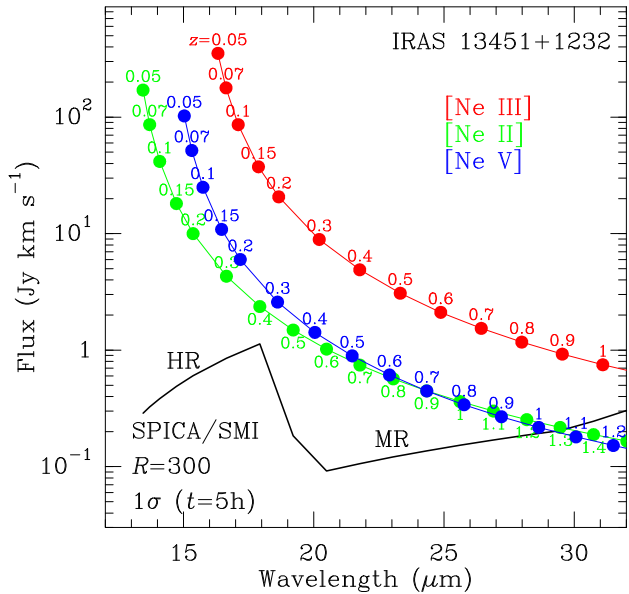}
\caption{Predicted integrated flux of the [Ne {\sc iii}]15.5$\mu$m (red), 
[Ne {\sc ii}]12.8$\mu$m (green), and [Ne {\sc v}]14.3$\mu$m (blue) lines, in
IRAS~13451+1232 at blueshifted velocities \citep[from $-3500$ \kms\ to $-500$
  \kms, from][]{spo09} as a function of redshift (small numbers) and observed
wavelength. The black curve shows the sensitivity ($1\sigma$) expected for 
{\it SPICA}/SMI HR and MR with a resolution of $R=300$ with 5 hours of
observing time. The ionized phase of the outflows in the mid-IR is detectable 
up to $z\sim0.7-1$.
} 
\label{ne}
\end{center}
\end{figure}

\subsection{Physical conditions of the molecular outflows using the CO band at
  $4.7$ $\mu$m in the mid-IR} 



Another crucial piece of information to study the physical conditions of the
molecular outflows is the near-IR CO ro-vibrational transitions ($\Delta
v =1$, $\Delta J = \pm 1$, $\lambda \sim 4.7\mu$m). \cite{shi13} observed the
CO band in absorption toward the nucleus of IRAS~08572+3915, clearly showing
that the main component of the absorption is blueshifted and thus probes the
outflow from the nucleus. The observed line absorption of $\sim50$\% of the
continuum indicates that the outflowing CO is covering a large fraction of the
mid-IR continuum, providing a sensitive probe of molecular feedback close to
the central engine. Moreover, since many lines at various excitation 
levels can be observed simultaneously, \cite{shi13} could estimate the gas
temperature of the outflowing gas, and hence the column density of the CO gas
with very little uncertainty.   

We can extend this work to the high-redshift universe with {\it SPICA}/SMI and
the James Webb Space Telescope ({\it JWST}, see also
  \S\ref{sec:synergies}). 
Figure~\ref{coband} shows the estimated flux of ULIRGs at the
wavelength around the CO feature together with the expected sensitivity of
{\it SPICA}/SMI/HR. A spectral resolution of $R=5000$ is required to avoid
blending of adjacent CO lines and thus to correctly place the continuum. With
the spectral coverage of $12-18$ $\mu$m, which corresponds to a redshift of 
$z= 1.5-3$ for the CO band, {\it SPICA}/SMI can observe the band toward
luminous ULIRGs with enough S/N ratio for the absorption study. 
Additional potential probes of molecular
outflows in the mid-IR include the bands of H$_2$O ($6.3$ $\mu$m), HCN (14
$\mu$m), CO$_2$ (15 $\mu$m), and C$_2$H$_2$ (13.7 $\mu$m).

\subsection{The ionized phase of outflows: fine-structure lines of Ne ions in
  the mid-IR and the [C {\sc ii}]157$\mu$m line} 
\label{sec:ion}


With {\it Spitzer}/IRS, blueshifted emission in the [Ne {\sc iii}]15.5$\mu$m, 
[Ne {\sc ii}]12.8$\mu$m, and [Ne {\sc v}]14.3$\mu$m fine-structure lines
was found in $\sim30$\% of local ULIRGs, most of them classified as AGNs
\citep{spo09,spo09b}. In some sources, 
the wings were detected up to $-3500$ \kms, with indications of higher
blueshift with increasing ionization state of the gas. The high ionization
potential of Ne$^{3+}$, $\approx97$ eV, ensures that the [Ne {\sc v}] line
can only be produced in gas directly irradiated by an AGN, thus unambiguously
probing the ionized phase of AGN feedback at wavelengths where extinction is
less severe than in the UV, optical, or near-IR 
Since the high velocity Ne gas is only seen in sources with weak silicate
absorption \citep[the lower branch of the fork diagram in][]{spo07},
indicating a very low dust column to the nucleus, a direct 
comparison of sources with molecular outflows and outflows in the high
ionization gas might probe different evolutionary states \citep{spo13}.


We use IRAS~13451+1232, where the three Ne lines are detected with
blueshifted emission up to at least $-3000$ \kms\ \citep{spo09}, as a local
template to compare in Fig.~\ref{ne} the expected fluxes for the blueshifted
emission as a function of redshift 
with sensitivity expectations for {\it SPICA}/SMI.
The three lines, at rest wavelengths of $12.8-15.5$ $\mu$m, shift with
increasing redshift from the SMI/HR band ($12-18$ $\mu$m,
$R_{\mathrm{nom}}=28000$) into the SMI/MR band ($18-36$ $\mu$m,
$R_{\mathrm{nom}}=1300-2300$); our sensitivity calculations correspond to a
smoothed resolution of $R=300$ (black curve in Fig.~\ref{ne}). Very
interestingly, strong outflows in the [Ne {\sc iii}] line will be
detected in 5h of observing time up to $z\sim1$, enabling direct comparison
with OH119 across the cosmic epoch of most pronounced decrease in star
formation rate.



In a sample of local far-IR bright galaxies, \cite{jan16} have reported on the
detection of broad wings in the [C {\sc ii}]157$\mu$m line, and have shown
that these wings are found in sources with high velocity outflowing gas as
seen in OH119. In addition, the outflow masses derived from OH and broad 
[C {\sc ii}] show a tentative $1:1$ relationship, suggesting that the
atomic and molecular gas phases of the outflow are connected
\citep{jan16}. Furthermore, the molecular outflow masses inferred from 
Na {\sc i} D \citep{rup13b}, CO \citep{cic14}, and OH (GA17), appear to show
similar agreement. The similarity of the derived masses using tracers that are
expected to arise from different phases or components, suggesting a phase
  change of the outflow material or high ionization rate of the
    outflowing molecular gas,
deserves a more statistically significant observational study. With  
the {\it SPICA}/SAFARI instrument, the [C {\sc ii}]157$\mu$m line would be
observed in the local Universe up to $z\approx0.45$.

\subsection{Inflows}
\label{sec:inflow}

\begin{figure*}
\begin{center}
\includegraphics[width=17cm]{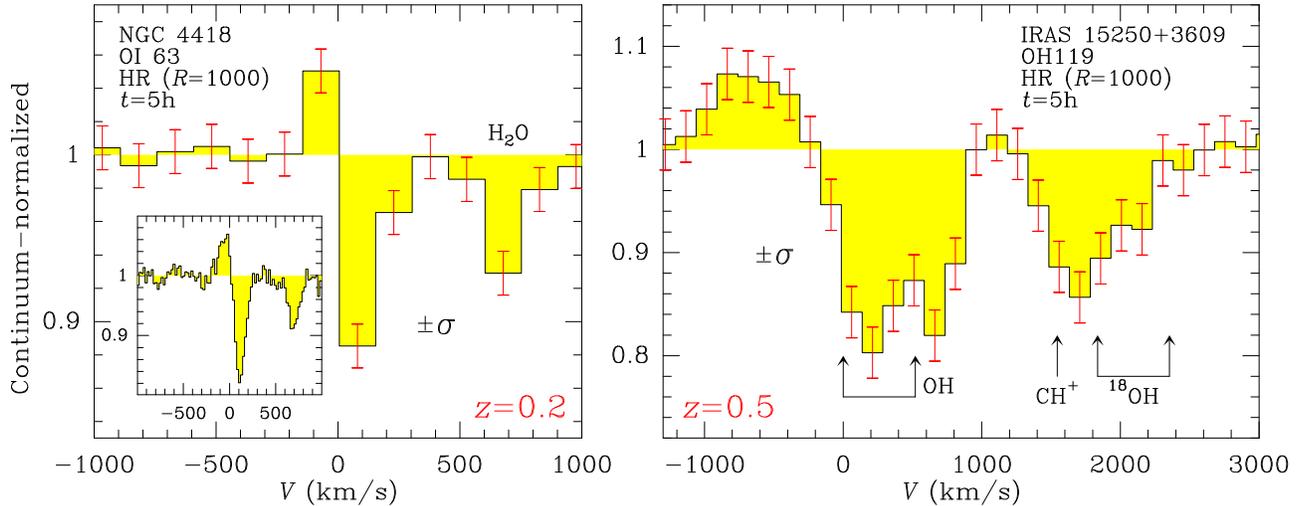}
\caption{{\it Left}: The [O {\sc i}]63 $\mu$m line at 63 $\mu$m in the LIRG
  NGC~4418 observed with {\it Herschel}/PACS, showing an inverse P-Cygni
  profile characteristic of galaxy-scale inflowing gas (GA12). The spectrum is
  smoothed to a resolution of $R=1000$, and errorbars indicate the expected
  $\pm\sigma$ uncertainties at $z=0.2$ for {\it SPICA}/SAFARI in HR mode with
  5 hours of observing time. The spectral feature at $\sim700$ \kms\ is a very
  excited H$_2$O line formed in the inner galaxy core. The insert shows the
  unsmoothed spectrum. {\it Right}: the OH119
  spectrum in IRAS~15250+3609, showing redshifted absorption. The strong
  feature at $1600-2200$ \kms\ is due to redshifted CH$^+$, because the red
  component of the $^{18}$OH doublet is not detected. Errorbars indicate the 
  expected $\pm\sigma$ uncertainties at $z=0.5$ for {\it SPICA}/SAFARI in HR
  mode with 5 hours of observing time. 
} 
\label{inflows}
\end{center}
\end{figure*}

In the local Universe, evidence for galaxy-scale inflows has come from 
{\it inverse} P-Cygni line shapes and redshifted absorption seen primarily in
both the [O {\sc i}]63 $\mu$m line and in the ground-state OH doublets. 
These inflows are usually spatially extended in comparison with the size of
the nuclei, thus probing the feeding of galaxy cores
\citep[GA12, GA17]{fal15,fal17}. 
While simulations predict that high-z galaxies are fed by relatively
  pristine gas, there is growing evidence that the immediate environments of
  high-z sources may be metal-rich \citep{pro14,emo16,nee17}, in which case
  {\it SPICA}/SAFARI can play an important role in studying 
  galaxy-scale inflows on significant cosmic timescales.
In local sources, the velocities associated with these motions are low,
and thus the HR mode of {\it SPICA}/SAFARI will be required to study the
  inflows.

Inflow signatures are found in {\it all} observed (in the far-IR) local LIRGs
(with $\sim(1-4)\times10^{11}$ \Lsun) that host a very compact and warm
nucleus (NGC~4418, Zw~049.057, Arp~299a, and IRAS~11506$-$3851), which 
may represent galaxies in an early stage of merging, or accreting
metal-rich gas from the intergalactic medium via efficient ``cold''
inflows. We show in Fig.~\ref{inflows} the inverse P-Cygni observed in the
[O~{\sc i}]63 $\mu$m line towards NGC~4418 ($L_{\mathrm{IR}}\sim10^{11}$
\Lsun, $z\approx0.007$) smoothed to a resolution of $R=1000$. The errorbars
indicate the $\pm\sigma$ uncertainty expected for {\it SPICA}/SAFARI in 5
hours of observing time at $z=0.2$. For a LIRG with
$L_{\mathrm{IR}}\sim4\times10^{11}$ \Lsun, this type of line shape would be
detected up to $z\approx0.4$. {\it SPICA}/SAFARI would thus obtain an
estimate of massive inflow rates ($\gtrsim10$ \Msun\ yr$^{-1}$) in LIRGs in
the local Universe and beyond. 

In the sample of sources observed with {\it Herschel}/PACS, there is one local
ULIRG with prominent redshifted absorption in the OH doublets, IRAS~15250+3609
($L_{\mathrm{IR}}\sim10^{12}$ \Lsun, GA17), probably indicating massive
funneling of gas toward the central galaxy core as a consequence of merging.
The spectrum of HCO$^+$ 3-2 at submillimeter wavelengths also shows a
redshifted absorption feature, but the profile is dominated by redshifted
emission and blueshifted absorption of the continuum characteristic of an
outflow \citep{ima16}\footnote{While the far-IR is mostly sensitive to the
  gas in front of the continuum source, blocking to some extent the emission
  from behind, the submillimeter traces the gas in front and behind the
  continuum more evenly, which may explain the observed differences.}. 
The redshifted OH119 absorption in this source is shown with $R=1000$ in
Fig.~\ref{inflows}. The emission feature at negative velocities is expected to
be associated with outflowing gas, but the redshifted absorption represents a
massive inflow. The errorbars illustrate that these type of profiles in
ULIRGs can be traced to $z\gtrsim0.5$ with SAFARI. The
absorption feature at $1600-2200$ \kms, almost as strong as OH119, appears to
be primarily due to redshifted CH$^+$, suggesting intense dissipation of
mechanical energy \citep{falg15}.

\subsection{Constraining SEDs from lines}
\label{sec:seds}


Decomposition of the SED of galaxies into AGN and SB contributions is a key
aspect of studies of the corresponding luminosity functions based on 
multi-wavelength photometric data-sets \citep[e.g.][]{gru13,gru16,del14}. 
While the fractional contributions to the luminosities by the AGN and the SB
are in most cases very well defined, there are cases where it may be relevant
to distinguish between nuclear and extended star formation, or where the
columns towards the nuclear regions are not well constrained with photometric
means, or where exceptionally high gas column densities may partially mask the
AGN. In these cases, observations of absorption, radiatively pumped molecular
lines in the far-IR can help resolve these ambiguities, as $T_{\mathrm{dust}}$
and the continuum optical depth are constrained from line modeling
(e.g. GA15).  The continuum emitted by these far-IR optically thick, warm
components is usually blended with and diluted within the 
contribution from colder dust, and hence the only way to dissentangle the cold
($\lesssim45$ K) and warm ($\gtrsim65$ K) components is through lines. 
The technique has been applied to local (U)LIRGs, enabling the
characterization of the SEDs associated with buried galaxy nuclei 
\citep[e.g.][GA12,GA17]{fal15,fal17}, and would be applied to high-$z$ sources
with the advent of {\it SPICA}.

Information from {\it emission} lines can also potentially be used as a prior
in SED fitting with the MCMC SED fitting codes which have already been
developed \citep[e.g. the SED Analysis Through Markov Chains,][]{joh13}.
\cite{efs14} also discussed whether the line information in IRAS~08572+3915 is
consistent with the idea that the ULIRG is powered by an AGN and a young
SB, which included calculations with CLOUDY \citep{fer13}.  

An interesting way of linking the feedback studies with SED fitting is to
try determine whether outflows are associated with galaxies in which the
SBs are relatively old, which is expected within the negative
feedback scenario of star formation quenching. As shown in earlier works, the
age of the SB can be an important parameter which determines the shape
of the IR spectrum and especially the prevalence of PAH features 
\citep[e.g.][]{efs00,row09}. High quality rest-frame mid-IR data of
galaxies will enable this kind of SED modeling, available with {\it JWST} and
also potentially with {\it SPICA} itself.

\subsection{Synergies with other missions}
\label{sec:synergies}


The study of outflows is among the strongest synergies between 
{\it SPICA} and {\it JWST}, {\it Athena}, ALMA, NOEMA, SKA, and
the E-ELT. Specific questions that could be addressed with the combined
capabilities of the four facilities  include: \\
1) What are the morphologies, masses, mass outflow rates, and kinetic 
energies of the observed outflows, and how do they relate to the
properties of the nucleus and the host? \\
2) What are the best tracers to obtain reliable estimates of the 
outflow properties? \\
3) How common are massive outflows at the peak epoch of AGN and 
galaxy assembly ($z = 1-3$)? \\
4) In which galaxy/AGN types do massive outflows occur? In 
which phase of the evolution of a galaxy and how long does the 
active feedback phase lasts? \\
5) How are molecular outflows linked to the ionized and atomic outflows and
how do they propagate?  


E-ELT/HARMONI will provide integral-field unit (IFU) observations in the
optical and near-IR ($0.47-2.45$ $\mu$m) enabling the mapping of host-galaxy
kinematics to distinguish between rotation signatures, irregular 
kinematics, and high-velocity components (i.e. potential 
signatures of outflowing gas) presumably up to redshifts of $\sim3$ 
\citep[e.g.][]{ken16}. Because E-ELT/HARMONI is a single IFU, 
obtaining constraints for a large sample of AGNs will 
be observationally expensive. The second generation multi-IFU 
E-ELT/MOS will enable observations of the kinematic components of 
several AGNs within a single field of view at $0.8-1.8$ $\mu$m, 
thus providing a catalogue of a statistically significant AGN 
sample. E-ELT spectroscopy will also provide additional diagnostics such 
as broad [Ne {\sc v}] emission, Al {\sc iii} and other UV absorption 
features, that will characterize ionized gas outflows at 
substantially higher redshifts than is currently possible.
The {\it SPICA} observatory would probe in $z<2$ galaxies 
the molecular counterpart, to verify whether the observed feedback is
accompanied by significant amounts of star-formation material. 

With the help of VLT-SINFONI IFU spectroscopy, \cite{can12}
obtained the [O {\sc iii}]$\lambda5007$ emission-line kinematics map of the 
luminous quasar 2QZJ002830.4-281706 at $z = 2.4$, that revealed 
a massive outflow on scales of several kpc. The detection of narrow 
H${\alpha}$ emission revealed non-uniformly distributed star 
formation in the host galaxy, with an SFR of $\sim 100$ M$_{\odot}$ 
yr$^{-1}$ strongly suppressed in the region where the highest 
outflow velocity and velocity dispersion are found. 
With an angular resolution and sampling of $0.01$ arcsec at 2 $\mu$m,  
E-ELT/HARMONI will be able to image the same field of 
view over $250\times250$ pixels with a spatial resolution of 80 pc.
Such observations will have a major impact on our understanding
of the effect of ionized gas winds on the star formation of the host galaxy 
and, given the extremely high $L_{\mathrm{AGN}}\approx10^{13}$ \Lsun,
{\it SPICA} would trace the clearing of the molecular gas reservoir from the
galactic disk (Section~\ref{sec:highz}).

In the foreseeable future, ALMA and NOEMA will continue to lead the way at 
(sub)millimeter wavelengths, providing maps of the 
cold gas components in CO and also denser tracers such as 
HCN, HCO$^+$, and CS. Given the long observing time required for
  obtaining spatially resolved observations of line wings at high redshifts,
and given that {\it SPICA}/SAFARI would enable the observation of $\sim1000$
galaxies in the LR mode at $z=0.5-3.5$, a possible strategy could consist of 
{\it SPICA} providing samples of galaxies with massive outflows to be
subsequently mapped with interferometers. Cross-check of outflows 
detected in the far-IR and (sub)millimeter would also be important to 
calibrate the inferred energetics.

{\it JWST}, which is scheduled to launch in late 2018, will provide
unprecedented imaging and spectroscopic capabilities in the near and
mid-infrared.  Specifically, the Near Infrared Spectrograph (NIRSpec), which 
will cover $0.6-5.3$ $\mu$m at $R\sim2700$, and the Mid
Infrared Instrument (MIRI), which will cover $4.9-28.8$ $\mu$m at
$R\sim1500-3500$, will both have Integral Field Unit (IFU) modes, enabling
high sensitivity, spatially-resolved spectra of star-forming galaxies and AGN.
As has been shown in studies of nearby AGN
\citep[e.g.,][]{mul16,rif16} and $z\sim1-2$ starbursts 
\citep[e.g.,][]{gen14,liv15,mie16},
IFUs are powerful tools to disentangle the excitation and kinematics of the
atomic and molecular gas in the nuclei of galaxies. 
At low redshifts, MIRI and NIRSpec will provide access to rest-frame near-IR
and mid-IR diagnostic features that can be used to separate out
photo-ionization and heating from shock excitation, on spatial scales that
range from tens to hundreds of parsecs.  
At high redshifts, the bright optical emission lines 
(e.g. H$\alpha$, [N {\sc ii}], [O {\sc i}], [S {\sc ii}]) will 
all pass into the NIRSpec bands, enabling traditional diagnostic diagrams
(e.g. the Baldwin, Phillips and Terlevich or BPT diagram) to be used to
identify shocked gas on scales of a few kpc. In the rest-frame near-IR, 
there are key diagnostics of shocks ([Fe {\sc ii}]] at 1.25 and 1.64 $\mu$m) 
and warm ($500-1000$ K) molecular gas (the $1-0$ S(1) H$_2$ line at 2.21
$\mu$m), which can be directly compared to the H {\sc ii}-region emission 
from hot, young stars as traced by the Paschen and Bracket series
recombination lines, and the PDR emission as traced by the 3.3
PAH emission feature. The mid-IR provides access to a suite of PAH
features 
to probe grain ionization and size, which can be altered by fast shocks in
outflowing winds \citep[e.g.,][]{bei15}. In addition, 
the H$_2$ S(0)$-$S(7) lines, and bright fine-structure cooling lines 
provide a sensitive probe of the warm
($100-500$ K) molecular gas and the ionization and density of the atomic ISM,
respectively.  Slow shocks can heat the molecular gas providing enhanced H$_2$
emission, and the large range in ionization potential and critical density 
of the fine structure lines can be used to not only detect buried AGN 
\citep[see][this issue]{spi17}, but also to uncover the presence of 
shocks \citep{all08,ina13}. A number of molecular
bands in the mid-IR (CO, H$_2$O, HCN, C$_2$H$_2$, and CO$_2$) can also provide
a unique view of the structure of the innermost phase of molecular outflows.

The Square Kilometre Array \citep[SKA, e.g.][]{mor15}, with early science
planned for 2020 and fully operational in 2030, will provide
HI 21 cm absorption surveys with unprecedented sensitivity, enabling the
detection of the atomic phase of outflows up to $z\sim3$ \citep{mor15}. 
Galaxy feedback processes at high redshifts would be studied in both the
atomic and molecular phases with SKA and {\it SPICA}. In addition, inflows 
are potentially detectable in the HI 21 cm with SKA 
\citep[as observed in the local NGC~4418,][]{cos13}, enabling the study of
their metallicity when compared with {\it SPICA} observations of the 
[O~{\sc i}]63 $\mu$m line and the OH doublets (Section~\ref{sec:inflow}).

The observing plans for {\it Athena} currently envisage an ambitious 
Wide Field Imager \citep[WFI,][]{mei16} survey including medium and deep
pointings which would allow the 
detection of thousands of (mostly mildly) Compton-thick (CT) AGN up to a
redshift of $3-4$ and their recognition as such up to a redshift of 3
\citep{car14}, characterising a few tens of those up to $z\sim3$. In addition,
it would also uncover X-ray nuclear UFOs in a few thousand type 1 AGN up to
$z\sim 3$ \citep{car15}. This is a rich harvest of potential targets for
detailed SPICA follow-up spectroscopy: CT sources could be targeted to
ascertain the putative relationship between heavy obscuration and molecular
outflows (Section~\ref{sec:buried}) comparing the incidence of these features
at different obscuration levels; {\it SPICA} could also observe populations of
type 1 AGN with and without UFOs, looking for the incidence of molecular
outflows in both populations, to investigate the connection between the energy
injected in the circumnuclear region by the AGN and the galaxy-wide outflows
that provide an effective feedback (Sections~\ref{sec:highz} and 
\ref{sec:h2ooh+}). 

In addition, {\it Athena} will also obtain high resolution X-IFU \citep{bar16}
spectroscopy of samples of AGNs out to $z\sim 2$ showing 
moderately ionised outflows, to measure the mechanical energy associated to
those outflows \citep{car15}. X-IFU spectroscopy will also be carried
out on nearby AGN outflows, to measure their kinetic energy and to understand
how the outflows are launched, also probing the interaction of winds from AGN
and star-formation with their surroundings \citep{pon15}.  This would
help understand how BHs quench their own mass reservoir and even how the 
$M - \sigma$  relation is established. At the same time, {\it SPICA} would
observe the molecular phase, and the putative relationship between nuclear
(parsec-scale) AGN winds and the more extended molecular outflows
\citep{tom15,fer15} would be revealed on the basis of significant galaxy
samples.

\section{Conclusions}

A breakthrough in our knowledge of molecular outflows through cosmic
time is expected with the IR observatory {\it SPICA}. 
We have shown in this paper that the unprecedented sensitivities predicted for
the SAFARI instrument would enable the detection and accurate description of
outflows in OH up to $z\sim1$, the limiting redshift for the observation of
the most sensitive OH doublet at 119 $\mu$m. It is in this epoch
when the strongest drop in IR luminosity density and star formation takes
place, and {\it SPICA} would check the role of massive outflows in star
formation quenching. In addition, bright sources will
also be detected in the more optically thin OH transition at 79 $\mu$m up to
$z\sim1.5-2$, i.e. the last $\sim10$ Gyr of the Universe. 
For the first time it will be possible to study molecular outflows in the
  infrared in massive, main sequence galaxies near their peak of activity,
  testing the relationship between feedback and the shape of the MS. 
We have shown predictions for lines of H$_2$O and OH$^+$ in the far-IR,
and also for the CO band at 4.7 $\mu$m and the strongest lines of ionized Ne
in the mid-IR. Inflowing gas associated with mergers or accretion of
metal-rich intergalactic gas will also be an exciting topic of study with
SPICA, through spectroscopic observations of the [O {\sc i}]63$\mu$m and OH
transitions. Finally, we have discussed the most important synergies that 
{\it SPICA} will have with other current and future observatories in the
  study of feedback and galactic evolution.

\begin{acknowledgements}
This paper is dedicated to the memory of Bruce Swinyard,
who initiated the {\it SPICA} project in Europe, but 
unfortunately died on 22 May 2015 at the age of 52. He was 
ISO-LWS calibration scientist, {\it Herschel}/SPIRE instrument 
scientist, first European PI of {\it SPICA} and first design lead 
of SAFARI.

We thank Corentin Schreiber and David Elbaz for providing us with the SEDs of
MS galaxies in electronic form \citep[reported in][]{sch15}.
EGA is a Research Associate at the Harvard-Smithsonian
Center for Astrophysics, and thanks the Spanish 
Ministerio de Econom\'{\i}a y Competitividad for support under projects
FIS2012-39162-C06-01 and  ESP2015-65597-C4-1-R. EGA and JF acknowledge support
under NASA grant ADAP NNX15AE56G. Basic research in IR astronomy at NRL
is funded by the US ONR. FJC acknowledges financial support through grant
AYA2015-64346-C2-1-P (MINECO/FEDER). 
This research has made use of NASA's Astrophysics Data System
(ADS) and of GILDAS software (http://www.iram.fr/IRAMFR/GILDAS).
\end{acknowledgements}

\begin{appendix}

\section{Scaling flux densities with redshift and sensitivities}
\label{appa}

We use the spectra observed by {\it Herschel}/PACS in some templates
and scale the flux densities with redshift $z$ according to
$F_{\nu,\mathrm{Jy}}\propto (1+z) / D_L^2$, where $D_L$ is the luminosity
distance. We use for $D_L$ a flat Universe with 
$H_0=69.6$ km s$^{-1}$ Mpc$^{-1}$ and $\Omega_{\mathrm{M}}=0.286$ \citep{ben14}. 

\begin{figure}
\begin{center}
\includegraphics[width=8cm]{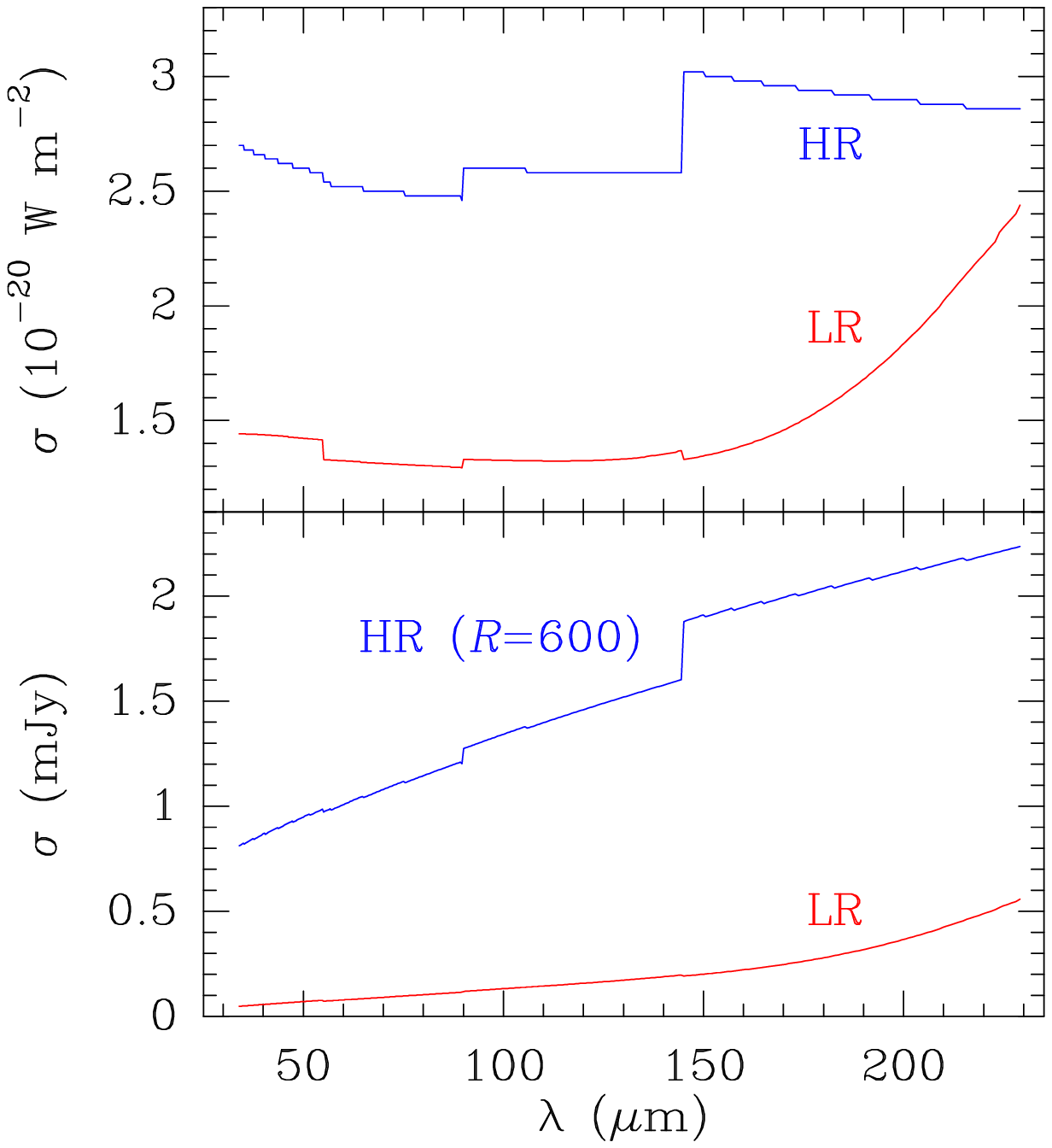}
\caption{{\it Upper}: Expected {\it SPICA}/SAFARI 1$\sigma$-1h 
  sensitivities in low-resolution (LR, red) and high-resolution (HR, blue)
  modes, at the nominal instrumental resolution \citep{spi17,roe17}. 
  {\it Lower}: {\it SPICA}/SAFARI
  1$\sigma$-1h sensitivities for flux densities in low-resolution and
  high-resolution modes, the latter smoothed to $R=600$ (see
  eq.~\ref{eq:sens}).  
} 
\label{sensit}
\end{center}
\end{figure}

Figure~\ref{sensit} shows the estimated 1$\sigma$-1h sensitivities currently
expected {\it SPICA}/SAFARI \citep{spi17,gru17} in low-resolution (LR) and
high-resolution (HR) modes, at their nominal resolution 
($R_{\mathrm{nom}}=300$ for LR and 
$R_{\mathrm{nom}}=2\times10^3\times200\mu m/\lambda$ 
for HR, where $\lambda$ is the observed wavelength). The corresponding
sensitivities for the flux densities are given by
\begin{equation}
\sigma_{\mathrm{mJy}} = 10^{29} \times
\frac{\sigma_{\mathrm{W/m^2}}\lambda_{\mathrm{\mu m}}}{c_{\mathrm{\mu m/s}}} 
 \times \sqrt{\frac{R_{\mathrm{nom}}R}{t_{\mathrm{h}}}},
\label{eq:sens}
\end{equation}
where the subscripts indicate units, $t$ is the observing time, and 
$R\leq R_{\mathrm{nom}}$ is the desired spectral resolution. Fig.~\ref{sensit} 
shows the calculated $\sigma_{\mathrm{mJy}}$ applied to both the LR and HR
modes (the latter smoothed to $R=600$) for $t_{\mathrm{h}}=1$. 
For the LR mode with $R=R_{\mathrm{nom}}$, $t_{\mathrm{h}}=2$, 
$\lambda_{\mathrm{\mu m}}=200$ (the wavelength of the OH79 doublet at $z=1.5$),
and $\sigma_{\mathrm{W/m^2}}=1.8\times10^{-20}$ (Fig.~\ref{sensit}),
$\sigma_{\mathrm{mJy}}\approx0.25$, or $\sigma_{\mathrm{norm}}\approx1$\% of
the continuum level for a continuum source of $F_{\nu}=24$ mJy (i.e. Mrk~231
scaled at $z=1.5$, see Fig.~\ref{oh79_lr}).
For the HR mode at $\lambda_{\mathrm{\mu m}}=226$
(the wavelength of the OH119 doublet at $z=0.9$, with
$R_{\mathrm{nom}}=1.77\times10^3$), eq.~(\ref{eq:sens}) with 
$R=600$, $t_{\mathrm{h}}=4$, and $\sigma_{\mathrm{W/m^2}}=2.9\times10^{-20}$
(Fig.~\ref{sensit}) yields $\sigma_{\mathrm{mJy}}\approx1.1$, or
$\sigma_{\mathrm{norm}}\approx2.5$\% of the continuum level for a continuum
source of $F_{\nu}=45$ mJy (i.e. Mrk~231 
scaled at $z=0.9$, see Fig.~\ref{oh119_hr}).

The 1$\sigma$ uncertainties of the line fluxes integrated from
$v_{\mathrm{min}}$ to $v_{\mathrm{max}}$ (black lines in the left panels of
Figs.~\ref{oh79_lr} and \ref{oh119_hr}) are calculated according to
$\sigma_{\mathrm{Jy\,km/s}}=0.5\times 10^{-3} \Delta v_{\mathrm{km/s}}
\sqrt{N_{\mathrm{spec}}}\, \sigma_{\mathrm{mJy}}$, where 
$\Delta v_{\mathrm{km/s}}$ is the spectral resolution of the (smoothed) 
spectrum, and $N_{\mathrm{spec}}=2\times(v_{\mathrm{max}}-v_{\mathrm{min}})/\Delta
v_{\mathrm{km/s}}$ is the number of spectral points in the velocity interval
(assuming a sampling of 2 points per resolution element).

\end{appendix}

\bibliographystyle{apj}

\nocite*{}
\bibliographystyle{pasa-mnras}
\bibliography{1r_lamboo_notes}

\end{document}